%% file: main.tex
\documentclass[preprint,12pt,authoryear]{elsarticle}

\input{preamble}
\journal{}
\begin{document}
 
	\begin{frontmatter}
 
		\title{Deficit-Driven Stratified Acquisition for Building
			STRATA-S12: An Open SAR–Optical Corpus for
			Self-Supervised Learning}
 
		\author[a]{Christian Sanchez-Leon}
		\ead{christian.sanchez.research@tutamail.com}

		\author[a]{Dongming Chen\corref{cor1}}
		\ead{chendm@mail.neu.edu.cn}
 
		\author[a]{Dongqi Wang}
 
		\cortext[cor1]{Corresponding author}
 
	\affiliation[a]{organization={Software College, Northeastern University},
		addressline={No. 195 Chuangxin Road, Hunnan District},
		city={Shenyang},
		postcode={110819},
		country={China}}

		\begin{abstract}
			Building a co-registered SAR--optical corpus for self-supervised Earth Observation is above all an acquisition problem, yet the main multi-modal datasets are collected opportunistically, with no per-stratum quota control, so the regions and seasons that cloud cover hits hardest end up under-sampled. We present \textbf{Deficit-Driven Stratified Acquisition (DDSA)}, an algorithm that casts the construction of a Sentinel-1/Sentinel-2 corpus as a constrained maximum-coverage problem over a land-cover-driven, hemisphere-corrected (class $\times$ season $\times$ region) stratification of target countries on all six inhabited continents, and solves it \emph{online}: acquisition is ordered by marginal gain against a per-stratum deficit that the quality gates and a visual review of every accepted pair keep updating during the run. Three results follow from measuring the algorithm on its own product. First, stratum balance is a property of the deficit-driven ordering rather than of run length: replayed over all {\SimTotalAttempts} real acquisition attempts, failures included, it populates all {\AlgoStratK} strata within {\SimFullDeficit}\% of the attempts, against {\SimFullRoundRobin}\% for uniform round-robin effort and {\SimFullRandom}\% for a random order. Second, whether a rejected attempt can be recovered is better predicted by how long its cause persists than by the type of defect: cloud rejections are recovered at $\eta={\AlgoEtaCloudsPct\%}$ against ${\AlgoEtaNodataPct\%}$ for \texttt{nodata}, consistent with clouds persisting through a season. Third, the repair loop supplies {\AlgoRepairSharePct\%} of the released pairs at no measurable quality cost, with {\AlgoQCSurvivalPct\%} surviving manual review, but it adds yield rather than balance. The quality guarantee has a stated price: {\AlgoNAttempts} acquisition attempts produced {\AlgoNPairs} released pairs. We release the corpus, \corpus, with per-pair metadata sidecars aligned with the STAC and PDS4 standards and the FAIR principles, and a pre-registered linear probe confirms it is fit for self-supervised pre-training.
		\end{abstract}
	
 
	\end{frontmatter}

	\input{sections/01_introduction}
	\FloatBarrier
	\input{sections/02_methods}
	\FloatBarrier
	\input{sections/03_implementation}
	\FloatBarrier
	\input{sections/04_corpus}
	\FloatBarrier
	\input{sections/05_validation}
	\FloatBarrier
	\input{sections/06_discussion}
	\FloatBarrier
	\input{sections/07_conclusions}
	\FloatBarrier
	\input{sections/08_backmatter}
	\FloatBarrier
	\bibliographystyle{elsarticle-harv}
	
	\bibliography{references}
	
\end{document}

%% file: preamble.tex
\usepackage{amsmath,amssymb}
\usepackage{graphicx}
\usepackage{subcaption}
\usepackage{booktabs}
\usepackage{hyperref}
\hypersetup{
	pdftitle={Deficit-Driven Stratified Acquisition for Building STRATA-S12: An Open SAR-Optical Corpus for Self-Supervised Learning},
	pdfauthor={Christian Sanchez-Leon, Dongming Chen, Dongqi Wang}
}
\usepackage[ruled,vlined,linesnumbered]{algorithm2e}
\usepackage{xcolor}
\usepackage[expansion=false]{microtype}
\usepackage{enumitem}
\usepackage{tikz}
\usepackage{array}
\usepackage{amsthm}
\usepackage{tabularx}
\newtheorem{proposition}{Proposition}
\usetikzlibrary{positioning,fit,backgrounds,arrows.meta,calc,shapes.geometric}
\setlist{nosep}

\hypersetup{colorlinks=true, allcolors=blue!55!black}

\newcommand{\phasebar}[1]{\BlankLine\tcc*[l]{\color{blue!60!black}\sffamily #1}}
\usepackage{siunitx}
\usepackage[skip=6pt]{caption}
\IfFileExists{reben/vars/probe_vars.tex}{\input{reben/vars/probe_vars}}{}

\IfFileExists{algo_evidence/statsDB/algo_numbers.tex}{\input{algo_evidence/statsDB/algo_numbers}}{}
\IfFileExists{algo_evidence/sim/sim_numbers.tex}{\input{algo_evidence/sim/sim_numbers}}{}

\usepackage{xspace}
\newcommand{\corpus}{STRATA-S12\xspace}
\newcommand{\dlite}{D_{\mathrm{lite}}}

\usepackage{placeins}

%% file: reben/vars/probe_vars.tex

\newcommand{\ProbeMicroPreFull}{0.5552\,$\pm$\,0.0022}
\newcommand{\ProbeMicroRndFull}{0.5158\,$\pm$\,0.0068}
\newcommand{\ProbeMacroPreFull}{0.3366\,$\pm$\,0.0004}
\newcommand{\ProbeMacroRndFull}{0.2920\,$\pm$\,0.0046}

\newcommand{\ProbeMacroDelta}{0.0446}

\newcommand{\ProbeCriterionMargin}{0.0304}
\newcommand{\ProbeSeeds}{3}
\newcommand{\ProbeClassesValid}{19}
\newcommand{\ProbeClassesTotal}{19}
\newcommand{\ProbeMaxTrain}{40{,}000}
\newcommand{\ProbeTrainSeed}{42}
\newcommand{\ProbeEncoderEpoch}{100}

%% file: algo_evidence/statsDB/algo_numbers.tex
\newcommand{\AlgoNPairs}{740{,}746}

\newcommand{\AlgoNAttempts}{2{,}928{,}370}
\newcommand{\AlgoNRejected}{2{,}079{,}035}
\newcommand{\AlgoNWithdrawn}{108{,}589}
\newcommand{\AlgoNCorruptPairs}{101}

\newcommand{\AlgoNRecovered}{292{,}432}
\newcommand{\AlgoRepairSharePct}{39.5}
\newcommand{\AlgoQCSurvivalPct}{89.8}
\newcommand{\AlgoEtaCloudsPct}{18.8}
\newcommand{\AlgoEtaNodataPct}{41.4}
\newcommand{\AlgoEtaHasInfPct}{13.9}

\newcommand{\AlgoDecPairs}{41{,}403}
\newcommand{\AlgoDecPct}{5.6}

\newcommand{\AlgoEtaNorthAmericaPct}{26.4}
\newcommand{\AlgoEtaEuropePct}{47.6}
\newcommand{\AlgoGapFirstPass}{1.32}
\newcommand{\AlgoGapRecovered}{1.29}
\newcommand{\AlgoPurityFirstPass}{0.698}
\newcommand{\AlgoPurityRecovered}{0.716}

\newcommand{\AlgoNTiles}{473{,}786}
\newcommand{\AlgoNRasterFiles}{1{,}481{,}492}
\newcommand{\AlgoStratK}{192}
\newcommand{\AlgoStratMean}{3{,}858}
\newcommand{\AlgoStratMin}{7}
\newcommand{\AlgoStratMax}{21{,}327}
\newcommand{\AlgoStratAboveHundred}{184}
\newcommand{\AlgoStratSparse}{8}
\newcommand{\AlgoNEtaRec}{562{,}747}
\newcommand{\AlgoNEtaExh}{912{,}290}
\newcommand{\AlgoNEtaPending}{603{,}998}
\newcommand{\AlgoEtaPct}{38.2}
\newcommand{\AlgoNLedgerSuccess}{562{,}747}

\newcommand{\AlgoNManual}{108{,}576}
\newcommand{\AlgoNManualTiles}{96{,}355}
\newcommand{\AlgoNRejEdge}{318}
\newcommand{\AlgoNRetriedEdge}{65}

%% file: algo_evidence/sim/sim_numbers.tex
\newcommand{\SimTotalAttempts}{\num{2928370}}
\newcommand{\SimPSuccessMin}{0.2}
\newcommand{\SimPSuccessMax}{55.7}
\newcommand{\SimFullDeficit}{0.05}
\newcommand{\SimFullRoundRobin}{4.03}
\newcommand{\SimFullRandom}{20.5}
\newcommand{\SimFullRandomMin}{6.9}
\newcommand{\SimFullRandomMax}{52.9}
\newcommand{\SimDeficitEntropyFive}{0.996}
\newcommand{\SimRoundRobinEntropyFive}{0.980}
\newcommand{\SimRandomEntropyFive}{0.904}

\newcommand{\SimDeficitPairsFive}{\num{32056}}
\newcommand{\SimRandomPairsFive}{\num{37025}}
\newcommand{\SimYieldCostFive}{13}
\newcommand{\SimYieldCostTwentyFive}{6}

\newcommand{\SimFullDeficitMin}{0.03}
\newcommand{\SimFullDeficitMax}{0.10}
\newcommand{\SimFullDeficitAttempts}{\num{1609}}
\newcommand{\SimFullRoundRobinMin}{0.5}
\newcommand{\SimFullRoundRobinMax}{11.5}
\newcommand{\SimRoundRobinPairsFive}{\num{40960}}
\newcommand{\SimYieldCostRRFive}{22}

%% file: sections/01_introduction.tex
\section{Introduction}
\label{sec:intro}

Earth Observation (EO) imagery from missions such as Sentinel-1 \citep{torres2012sentinel1} and Sentinel-2 \citep{drusch2012sentinel2} has reached petabyte scale, far outpacing the human capacity to label it. Self-supervised learning (SSL) has emerged as a practical answer~\citep{wang2022sslreview}: by pre-training on large unlabelled corpora and then transferring to specific tasks with a fraction of the labels, SSL models reduce the annotation bottleneck that dominates supervised pipelines.

The effectiveness of any SSL model is, however, only as good as the corpus it is pre-trained on. In the multi-modal SAR--optical setting, the most widely used corpora --- SEN12MS \citep{schmitt2019sen12ms}, SSL4EO-S12 \citep{wang2023ssl4eo}, and to a lesser extent BigEarthNet \citep{sumbul2019bigearthnet} --- have been assembled \emph{opportunistically}: candidate scenes are pulled from publicly available archives and filtered under each corpus's own quality criteria, with little or no per-stratum quota control and no explicit mechanism for counteracting the geographic dropout introduced by non-uniform cloud cover, a bias the SSL4EO-S12 authors themselves acknowledge~\citep{wang2023ssl4eo}; Table~\ref{tab:features} summarises how these corpora differ from \corpus in content and acquisition design. The result is a well-documented \citep{torralba2011bias} but rarely-addressed problem: representational biases of the pre-training corpus propagate into the downstream model, a shift that domain-adaptation methods then attempt to correct~\citep{tuia2016domainadaptation}, and the spatial distribution of pre-training data has been shown to shape the behaviour of geospatial models~\citep{spatialdist2025}.

Counteracting this bias has recently received explicit attention. Several works sample a coordinate pool under a stratified scheme ---by continent or biome~\citep{spatialdist2025,nedungadi2024mmearth}, by land cover over a hemisphere split~\citep{presto2023}, or into attribute-balanced working subsets~\citep{akupara2025} --- while dense global gridding reduces bias through sheer coverage~\citep{majortom2024}. The same concern runs through recent large EO pre-training corpora and models, which pair many sensors, modalities or label types at scale~\citep{bastani2023satlas,geomeld2026,thor2026}. Stratification is also the established design in remote sensing for map accuracy assessment and area estimation, where strata are drawn from a map and samples are allocated across them under a probability design~\citep{stehman2009sampling,olofsson2014good}. DDSA borrows the stratification but uses it for a different purpose: its per-stratum quotas set coverage targets for a training corpus, not inclusion probabilities for an estimator, so the released corpus is balanced by design but is not a probability sample of the frame and is not meant to support design-based area estimates. DDSA differs in \emph{mechanism} rather than in goal: where these approaches stratify an \emph{offline} selection of available scenes, DDSA stratifies \emph{during} acquisition. It keeps a per-stratum deficit, orders the acquisition by marginal gain against it, and updates it whenever a quality gate or the human review removes a pair, so the strata that non-uniform cloud cover empties are moved back up the queue rather than left behind. To our knowledge, no existing SAR--optical corpus is acquired under such an online deficit, coupled with strict temporal pairing and quality gating for co-registered Sentinel-1/Sentinel-2 pairs.

This paper introduces \textbf{Deficit-Driven Stratified Acquisition (DDSA)}, an algorithm for the efficient, quality-controlled and coverage-maximising acquisition of co-registered Sentinel-1 / Sentinel-2 pairs, together with the open corpus it produces. Its sampling frame is defined by land cover rather than by population: candidate locations come from the WorldCover map of target countries on all six inhabited continents, not from the surroundings of cities, so the corpus is built as a general-purpose pre-training substrate rather than an urban-centred one. DDSA rests on three pillars:
\begin{itemize}
	\item \textbf{Download efficiency}: greedy maximum-coverage block planning (acquire-once / extract-all) over a stratification of the planet, minimising redundant Google Earth Engine downloads~\citep{gorelick2017gee}. This is efficiency in \emph{downloads} per block, not in the ratio of acquisition attempts to released pairs, which the quality gate deliberately trades away
	(Section~\ref{sec:datarecords}); 
	\item \textbf{Quality guarantee}: every released pair passes a strict temporal-pairing window and a multi-stage cloud, nodata and edge-quality cascade anchored on Sen2Cor scene-classification layers~\citep{mainknorn2017sen2cor};
	\item \textbf{Coverage control}: acquisition is ordered online by marginal gain against a per-stratum deficit, so every stratum is served before any one saturates; a human-in-the-loop, tile-level repair loop then refills attempts the gates rejected, adding yield at unchanged quality.
\end{itemize}

The greedy maximum-coverage rule at the core of the planning stage is a classical result~\citep{nemhauser1978submodular}, and we adopt it precisely because it is optimal: for the constrained maximum-coverage problem it solves (Section~\ref{sec:formulation}), no polynomial-time algorithm can guarantee a better coverage than its $(1-1/e)$ bound unless $\mathrm{P}=\mathrm{NP}$ \citep{feige1998threshold}. The contribution is therefore not the planner but what is built on it and what it allows us to measure. Specifically, this paper contributes:
\begin{enumerate}[label=(\roman*)]
	\item an online formulation of stratified acquisition, in which the greedy plan runs against a per-stratum deficit over a hemisphere-corrected (class $\times$ season $\times$ region) stratification, and the deficit is kept up to date by a strict pairing-and-gating cascade and a human-in-the-loop review (Section~\ref{sec:methods});
	\item evidence that stratum balance comes from this ordering and not from how far the run was taken: replayed over all {\SimTotalAttempts} acquisition attempts, failures included, deficit-driven ordering populates every stratum within {\SimFullDeficit}\% of the attempts, where uniform round-robin effort needs {\SimFullRoundRobin}\% and a random order {\SimFullRandom}\%, at the price of a lower yield per attempt (Section~\ref{sec:phasec-ordering});
	\item a per-reason measurement of recoverability over {\AlgoNAttempts} acquisition attempts, showing that the type of defect is a poor predictor of whether a retry succeeds, with the temporal persistence of its cause as the most parsimonious explanation (Section~\ref{sec:rejections});
	\item evidence that the repair loop supplies {\AlgoRepairSharePct\%} of the corpus at no measurable cost in quality, and that it adds yield rather than balance (Section~\ref{sec:algorithm-evidence}).
\end{enumerate}

The algorithm produces a single corpus released in two complementary modes: the \emph{full corpus} $D$, whose distribution reflects the true land-cover geography of the worked regions; and a derived \emph{reduced benchmark split} $\dlite$, in which every non-empty stratum is down-sampled to a common cap for controlled and resource-constrained use. Strata holding fewer pairs than the cap enter whole, so the split is capped rather than equalised. Each released pair carries a rich, standards-aligned JSON sidecar: its schema is aligned with STAC~1.0.0~\citep{stac2021spec}, informed by the NASA PDS4 information model~\citep{nasapds4}, and designed to satisfy the FAIR principles~\citep{wilkinson2016fair}.

We further provide a pre-registered technical validation showing, via frozen linear probing of an SSL backbone pre-trained on a representative subset of the corpus $D$ (with a random-initialisation baseline and multi-seed aggregation), that the released corpus is fit for its declared purpose of SSL representation learning in EO. We additionally characterise the corpus's geographic and class coverage and compare it, at the metadata level, against opportunistically assembled corpora (SSL4EO-S12, SEN12MS), quantifying the difference in how the two acquisition regimes distribute samples over a shared frame (Section~\ref{sec:coverage-comparison}). Systematic comparison of \emph{downstream SSL performance} across corpora and methods is deliberately left to follow-up work.

%% file: sections/02_methods.tex
\section{The DDSA Algorithm}
\label{sec:methods}

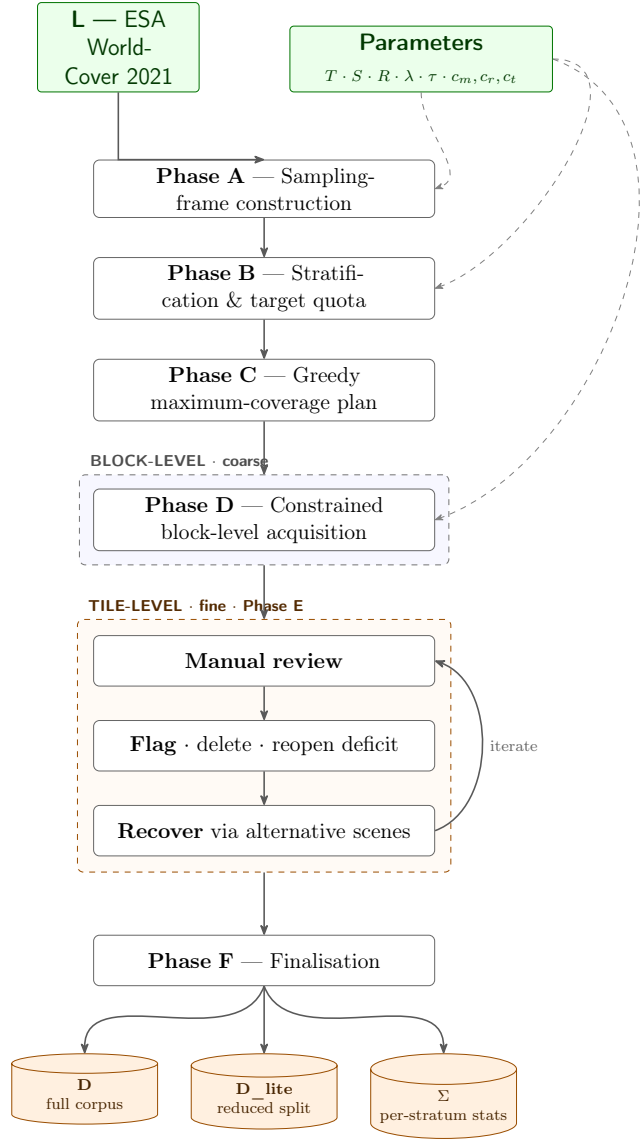
\begin{figure}[htbp]
	\centering
	\resizebox{!}{0.78\textheight}{%
		\begin{tikzpicture}[
			font=\sffamily,
			>={Stealth[round]},
			phase/.style ={rounded corners=3pt, draw=black!60, fill=white,
				align=center, inner sep=4pt, text width=58mm,
				minimum height=9mm, font=\small},
			inp/.style   ={rounded corners=2pt, draw=green!45!black, fill=green!8,
				text=green!22!black, align=center, inner sep=4pt,
				minimum height=9mm},
			outp/.style  ={cylinder, shape border rotate=90, aspect=0.22,
				draw=orange!60!black, fill=orange!12, text=orange!28!black,
				align=center, inner sep=3pt, font=\scriptsize,
				minimum width=26mm, minimum height=12mm},
			flow/.style  ={->, thick, black!65},
			param/.style ={->, dashed, black!50}
			]
			
			\node[phase] (a) {\textbf{Phase A} --- Sampling-frame construction};
			\node[phase, below=7mm of a] (b) {\textbf{Phase B} --- Stratification \& target quota};
			\node[phase, below=7mm of b] (c) {\textbf{Phase C} --- Greedy maximum-coverage plan};
			\node[phase, below=12mm of c] (d) {\textbf{Phase D} --- Constrained block-level acquisition};
			
			\node[phase, below=15mm of d] (e1) {\textbf{Manual review}};
			\node[phase, below=6mm of e1] (e2) {\textbf{Flag} $\cdot$ delete $\cdot$ reopen deficit};
			\node[phase, below=6mm of e2] (e3) {\textbf{Recover} via alternative scenes};
			
			\node[phase, below=14mm of e3] (f) {\textbf{Phase F} --- Finalisation};
			
			\node[inp, above=12mm of a.north, xshift=-26mm, text width=26mm] (l)
			{\textbf{L} --- ESA WorldCover 2021};
			\node[inp, above=12mm of a.north, xshift=28mm, text width=44mm] (p)
			{\textbf{Parameters}\\[1pt]{\scriptsize $T \cdot S \cdot R \cdot \lambda \cdot \tau \cdot c_m, c_r, c_t$}};
			
			\node[outp, below=12mm of f, xshift=-32mm] (dfull) {\textbf{D}\\ full corpus};
			\node[outp, below=12mm of f] (dbal) {\textbf{D\_lite}\\ reduced split};
			\node[outp, below=12mm of f, xshift=32mm] (sig) {\textbf{$\Sigma$}\\ per-stratum stats};
			
			\begin{scope}[on background layer]
				\node[draw=black!55, dashed, rounded corners=3pt, fill=blue!3,
				fit=(d), inner sep=7pt] (blk) {};
				\node[draw=orange!60!black, dashed, rounded corners=3pt, fill=orange!4,
				fit=(e1)(e2)(e3), inner sep=8pt] (tile) {};
			\end{scope}
			
			\node[font=\sffamily\bfseries\scriptsize, text=black!70,
			anchor=south west] at (blk.north west) {\,BLOCK-LEVEL $\cdot$ coarse};
			\node[font=\sffamily\bfseries\scriptsize, text=orange!32!black,
			anchor=south west] at (tile.north west) {\,TILE-LEVEL $\cdot$ fine $\cdot$ Phase E};
			
			\draw[flow] (l) -- (a.north -| l) -- (a.north);
			\draw[flow] (a) -- (b);
			\draw[flow] (b) -- (c);
			\draw[flow] (c) -- (blk);
			\draw[flow] (blk) -- (tile);
			\draw[flow] (tile) -- (f);
			
			\draw[flow] (e1) -- (e2);
			\draw[flow] (e2) -- (e3);
			\draw[flow] (e3.east) to[out=20,in=-20] node[right,font=\scriptsize,black!60]{iterate} (e1.east);
			
			\draw[flow] (f.south) to[out=-110,in=90] (dfull.north);
			\draw[flow] (f) -- (dbal);
			\draw[flow] (f.south) to[out=-70,in=90] (sig.north);
			
			\draw[param] (p.south) to[out=-90,in=20]  (a.east);
			\draw[param] (p.east)  to[out=0,in=20]    (b.east);
			\draw[param] (p.east)  to[out=-20,in=20]  (d.east);
			
		\end{tikzpicture}
	}
	\caption{DDSA algorithmic flow. The 10\,m ESA WorldCover map $L$ and a set of parameters (target $T$, strata $S,R$, sensor latitude limits $\lambda$, temporal tolerance $\tau$, cloud thresholds $c_m,c_r,c_t$) drive a six-phase pipeline. Phases~A--C build, stratify and order an acquisition plan. The block-level bulk acquisition (Phase~D) is followed by an iterated human-in-the-loop tile-level recovery (Phase~E), which re-opens the deficit of every pair it removes and refills rejected attempts with alternative scenes. Phase~F derives the full corpus $D$, the reduced benchmark split $\dlite$, and the per-stratum statistics report $\Sigma$.}
	\label{fig:flow}
\end{figure}

Figure~\ref{fig:flow} gives the high-level flow of the algorithm. Algorithms~\ref{alg:ddsa1} and~\ref{alg:ddsa2} formalise the procedure as two parts: planning (Phases~A--C, Algorithm~\ref{alg:ddsa1}) and acquisition with recovery (Phases~D--F, Algorithm~\ref{alg:ddsa2}). The remainder of this section first states the planning stage as a constrained submodular-maximisation problem (Section~\ref{sec:formulation}) and then describes each phase in prose.

\subsection{Problem formulation}
\label{sec:formulation}

The planning stage (Phases~A--C) is an instance of monotone submodular maximisation under a cardinality constraint, i.e.\ a maximum-coverage problem. Stating it as one makes both the choice of a greedy rule and the behaviour reported in Section~\ref{sec:phasec-ordering} --- that a partial run is already a covering one --- precise rather than incidental. 

Let the sampling frame be $F$ and let the strata be the product set $\Omega = C \times S \times R$, with $K = |\Omega| = \AlgoStratK$ in the current release (Section~\ref{sec:datarecords}). Cells are aggregated into blocks $b \in B$, and the count $\mathit{contrib}[b,s] \geq 0$ gives the tiles of $b$ that fall in stratum $s$ over all seasons (Phase~C). Each stratum carries a target quota $q$ (Phase~B), so the initial deficit is $d_0[s] = q$ for every $s$.

For a set of selected blocks $B' \subseteq B$, the coverage attained against the deficit caps is:
\begin{equation}
	f(B') \;=\; \sum_{s \in \Omega}
	\min\!\Big( d_0[s],\; \textstyle\sum_{b \in B'} \mathit{contrib}[b,s] \Big).
	\label{eq:coverage}
\end{equation}
Each summand is a concave non-decreasing function of a modular set function and is therefore submodular; a non-negative sum of submodular functions is submodular, so $f$ is monotone and submodular. Under a budget of $k$ blocks (one GEE download each), the planning objective is:
\begin{equation}
	\max_{B' \subseteq B,\; |B'| \leq k} f(B').
	\label{eq:maxcov}
\end{equation}

The maximum-coverage special case of~\eqref{eq:maxcov} is NP-hard, and no polynomial-time algorithm can guarantee a coverage better than a $(1-1/e)$ fraction of the optimum unless $\mathrm{P}=\mathrm{NP}$ \citep{feige1998threshold}. The greedy rule that adds, at each step, the block of largest marginal gain attains exactly this bound \citep{nemhauser1978submodular}. That marginal gain is:
\begin{equation}
	\Delta(b \mid B') \;=\; f(B' \cup \{b\}) - f(B')
	\;=\; \sum_{s \in \Omega} \min\!\big( \mathit{contrib}[b,s],\; d[s] \big),
	\label{eq:marginal}
\end{equation}
with residual deficit $d[s] = \max\!\big(0,\, d_0[s] - \sum_{b' \in B'} \mathit{contrib}[b',s]\big)$. The right-hand side of~\eqref{eq:marginal} is exactly the priority $\Pi(b)$ of Phase~C: the rule used here is greedy on the submodular objective~\eqref{eq:coverage}, and $\Pi(b)$ is its marginal gain rather than an ad-hoc score.

\begin{proposition}[Greedy coverage guarantee]
	\label{prop:greedy}
	The coverage $f$ of~\eqref{eq:coverage} is monotone and submodular. For any
	budget $k$, the greedy plan $B'_k$ obtained by iterating $\arg\max_{b} \Delta(b \mid B')$ satisfies $f(B'_k) \geq (1-1/e)\,\max_{|B'| \leq k} f(B')$, and no polynomial-time algorithm improves on the factor $(1-1/e)$ for the maximum-coverage special case unless $\mathrm{P}=\mathrm{NP}$.
\end{proposition}

Because $f$ has diminishing returns, the lazy re-evaluation of Algorithm~\ref{alg:ddsa1} --- re-scoring a block only when it surfaces at the top of the heap --- returns the same plan as full greedy at every step. The max-heap with on-pop re-scoring is thus an exact acceleration, not an approximation of it.

Proposition~\ref{prop:greedy} bounds the coverage of the \emph{plan}, not of the delivered corpus. Acquisition-time gates (regional and per-tile cloud, nodata, edge padding, and the inter-block overlap constraint of Phase~D) remove tiles the plan counted on, and non-uniform cloud removes them unevenly across $\Omega$. The realised coverage is $f$ evaluated over the tiles that survive Phase~D, which falls below the planned value wherever the sky was not clear --- the geographic dropout the corpus exists to counteract. DDSA handles that gap online rather than after the fact: the plan is executed in dynamic stratum-priority order, so every tile a gate removes leaves its stratum's deficit open and that stratum higher in the queue, and the deficit-repair loop described next retries the rejected attempts themselves.

The loop carries the residual deficit $d[\cdot]$ as state. A manual removal in Phase~E returns a unit to its stratum ($d[s] \leftarrow d[s]+1$) and excludes the offending source scene; a successful recovery consumes one ($d[s] \leftarrow d[s]-1$). Locations are visited in decreasing $d[s]$ and each is retried at most twice (Phase~E), so the iteration is bounded and terminates when no pair is flagged and every deficit is either cleared or backed by an exhausted candidate pool. What the loop does \emph{not} guarantee is $d[\cdot]=0$ at termination: a stratum whose season is overcast on every available date reaches the retry bound with $d[s]>0$, and the run is in any case stopped by the acquisition budget rather than by exhaustion of the frame (Section~\ref{sec:datarecords}). The share of retried attempts the loop refills is its recovery efficiency $\eta$, defined and measured per rejection reason in Section~\ref{sec:rejections} (Equation~\eqref{eq:recovery}).

The classical component of DDSA is thus the greedy plan, whose guarantee is already $(1-1/e)$-optimal and therefore not something a different heuristic could improve upon. What the algorithm adds is the online use of that plan: the per-stratum deficit is updated by every gate rejection, manual removal and recovery, and the acquisition order follows it. Section~\ref{sec:datarecords} measures what each part contributes. The deficit-driven ordering is what holds the $(\text{class}\times\text{season}\times\text{region})$ balance of the delivered corpus (Section~\ref{sec:phasec-ordering}); the repair loop adds yield at unchanged quality but does not, by itself, redistribute the corpus (Section~\ref{sec:coverage-comparison}).

\subsection{Phase A --- Sampling-frame construction}
A regular tiling of the ESA WorldCover 2021 map (product version v200; \citealp{zanaga2022worldcover}), delimited to a set of target countries, produces the candidate cells $t$. Centring the frame on the target countries makes their territory the geographic domain, which keeps the frame on mapped land and inland water while excluding open ocean by construction. Each cell is assigned its dominant land-cover class $c_t$ (the majority class over its pixels) and its continental region $g_t$ from Natural Earth admin-0 country boundaries~\citep{naturalearth}, each country mapped to its continent. Cells outside the per-sensor latitude window $|\mathrm{lat}_t| \leq \lambda$ are discarded. The remaining cells form the sampling frame $F$, which in the current release spans a configurable set of target countries across the six inhabited continents (Section~\ref{sec:datarecords}). The hemisphere-symmetric latitude filter and the WorldCover-derived class labels make the frame self-consistent across the worked regions.

\subsection{Phase B --- Stratification and target quota}
Strata are defined as the product of the land-cover classes present in $F$, the seasons $S$, and the continental regions $R$. The class axis is restricted to the eight WorldCover classes that occur across all sampled regions; classes absent from at least one region are dropped, since they would otherwise leave structurally empty strata. The hemisphere-aware season mapping ensures that, e.g., June--July--August is \emph{summer} in the Northern hemisphere and \emph{winter} in the Southern. Seasons follow the three-month meteorological windows (DJF, MAM, JJA, SON) rather than astronomical ones, and each window is named after the year in which it closes: the winter of year $Y$ runs from December of $Y-1$ to February of $Y$. Acquisitions dated December therefore belong to the following year's season, so a corpus collected over a single season cycle legitimately carries scenes from two calendar years. Together with the hemisphere correction this means the same month can take either label:
December is winter in the North and summer in the South, and both assignments appear in the corpus. Targets are set per land-cover class, so that acquisition stays balanced across classes rather than exhausting a tile-rich class before the others are filled; the per-class target $T$ is spread evenly over that class's strata, giving a per-stratum quota $q = \lfloor T/(|R|\cdot|S|) \rfloor$. This target is a balancing and stopping criterion, not a hard cap: the realised per-stratum counts are reported in $\Sigma$ (Section~\ref{sec:datarecords}).

\subsection{Phase C --- Greedy maximum-coverage block plan}
Cells are aggregated into rectangular acquisition blocks $b \in B$ (one GEE download per block; acquire-once / extract-all). For each block and each stratum $s$, the contribution $\mathit{contrib}[b,s]$ is the number of tiles of $b$ that fall in $s$ across all seasons. The greedy maximum-coverage heuristic \citep{nemhauser1978submodular} orders blocks by the priority $\Pi(b)$ of Section~\ref{sec:formulation} --- the marginal gain of the submodular coverage objective --- $\Pi(b)=\sum_{s}\min(\mathit{contrib}[b,s],\mathit{deficit}[s])$,
maintained over a max-heap with on-pop re-scoring to handle stale keys. Phase~C yields an ordered acquisition plan together with the residual deficit vector $\mathit{deficit}[\cdot]$. 

\subsection{Phase D --- Constrained block-level acquisition}
For each block of the plan, an optical scene is selected following the classic MODIS strategy~\citep{justice2002modis}: the \emph{most recent} Sentinel-2 Level-2A scene within the target season under the metadata cloud threshold $c_m$, refined by a regional Sen2Cor SCL cloud threshold $c_r$, and excluding scenes already known to have failed. A Sentinel-1 GRD scene is paired within a temporal window $|t(a)-t(o)| \leq \tau$. Each tile $p$ extracted from the block is then accepted only if it passes the full reject cascade: regional cloud $\leq c_r$, per-tile cloud $\leq c_t$, and no nodata or edge-padding violations. Land-cover purity is not an acceptance threshold; instead, each accepted tile is characterised by its dominant class, the purity of that class, and its top-three label composition (class id, name and pixel ratio), all taken from the WorldCover overlay and written to the sidecar. Mixed pixels are therefore described rather than discarded, and users can later keep homogeneous tiles,
work with mixed ones, or filter by any class present (Section~\ref{sec:datarecords}). In addition, a spatial-sampling-discipline constraint requires that the centre-to-centre distance from $p$ to any tile already accepted from a different block in the same season exceed $(1-r)\cdot \mathrm{footprint}_p$, with $r=0.30$, capping inter-block tile overlap at 30\% and preventing the redundant geographic
representation that block expansion or sub-block splitting could otherwise introduce. This is a purely \emph{geographic} overlap of two tile footprints in the same season, enforced automatically; it is distinct from the \emph{temporal} within-tile overlap caught by the Phase~E human review, where a single tile blends two acquisition dates. Accepted tiles increment the per-stratum counter $\mathit{counter}[s_p]$. Within each season up to two atomic Sentinel-1/Sentinel-2 pairs are acquired per location (the balanced setting; one is fastest, three is densest), and the three candidate months are visited in a fixed interleaved order (first, third, second), so that the pairs land in the first and third months rather than clustering in one; this temporal spread is one of the invariants re-checked by the post-acquisition audit (Section~\ref{sec:implementation}).

\subsection{Phase E --- Human-in-the-loop QC and tile-level recovery}
\begin{figure}[htbp]
	\centering
	\setlength{\tabcolsep}{2pt}
	\begin{tabular}{cc}
		\subcaptionbox{Small residual clouds / haze --- optical\label{fig:pe-a}}{%
			\includegraphics[width=0.46\linewidth]{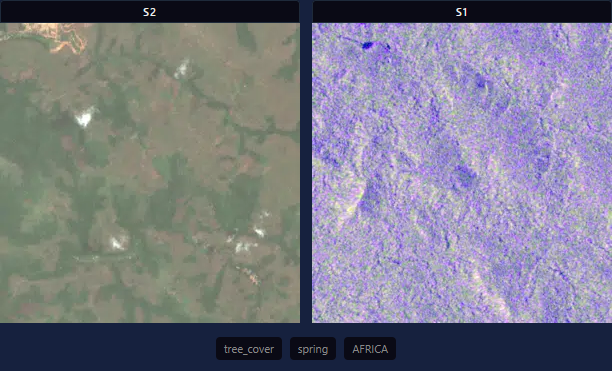}} &
		\subcaptionbox{Cloud shadow --- missed by the brightness gate\label{fig:pe-b}}{%
			\includegraphics[width=0.46\linewidth]{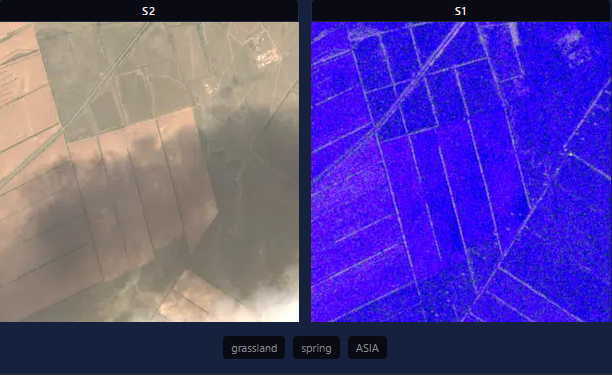}} \\[2pt]
		\subcaptionbox{Temporal seam --- two dates, in both sensors\label{fig:pe-c}}{%
			\includegraphics[width=0.46\linewidth]{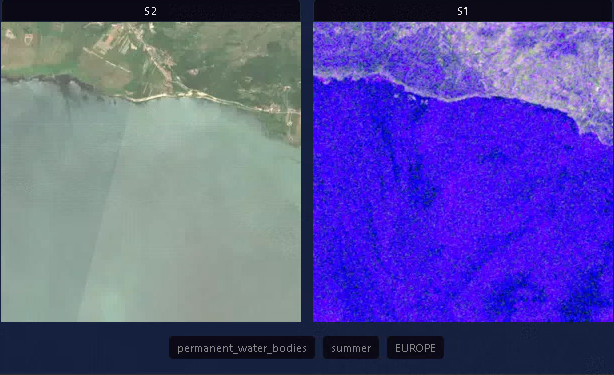}} &
		\subcaptionbox{Bright cross-shaped SAR return over metal (a ship)\label{fig:pe-d}}{%
			\includegraphics[width=0.46\linewidth]{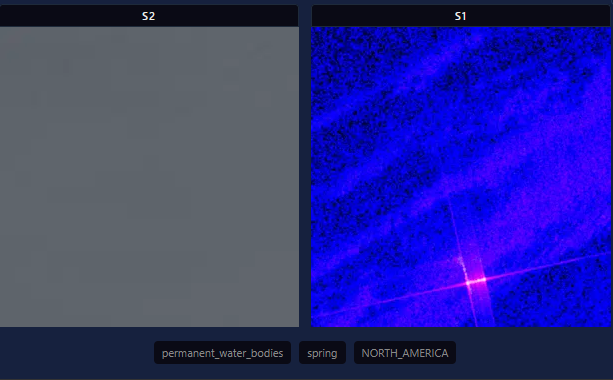}} \\[2pt]
		\multicolumn{2}{c}{%
			\subcaptionbox{Sentinel-2 inter-band offset --- a fast-moving aircraft split into colour fringes\label{fig:pe-e}}{%
				\includegraphics[width=0.46\linewidth]{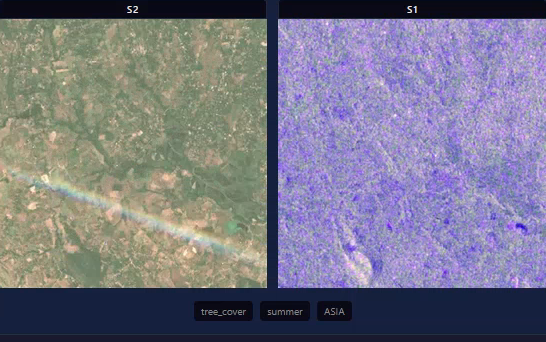}}} \\
	\end{tabular}
	\caption{Representative Phase-E manual removals, one per failure mode. In each 	panel the left half is the Sentinel-2 optical tile and the right half the Sentinel-1 SAR composite of the same pair. These are defects the automatic Phase-D gates cannot detect by construction; for the temporal seam~(c) the 	discontinuity appears co-registered in both the optical and the SAR channel, which a single-sensor artefact cannot produce. Dense fire-smoke removals also occur but are not shown. The complete list, with source-scene identifiers, is the Phase-E ledger released with the corpus.}
	\label{fig:phaseE-examples}
\end{figure}

Phase~D rejects tiles automatically, but residual cloud and subtle artefacts can still survive the gates. Phase~E iterates a bounded review/recovery loop, operating at \emph{tile} rather than \emph{block} granularity. Every pair the automatic gates accept is inspected visually before it can enter the corpus, and so is every replacement the recovery step brings back: each review round examines all pairs accepted since the previous one, so no pair is released without a human check. Review is carried out in a purpose-built local gallery that shows each pair side by side, the Sentinel-2 true-colour tile next to a Sentinel-1 false-colour composite of VV, VH and VV$-$VH in dB (Figure~\ref{fig:phaseE-examples}); the optical rendering switches to an adaptive stretch on hazy or over-exposed tiles, so that faint cloud and shadow stay visible, and the gallery does not indicate whether a pair came from the recovery step. The reviewer applies a fixed, closed set of criteria, each one targeting a failure mode that the automatic gates cannot detect by construction: 
(i) small residual clouds or haze that slip below the brightness threshold;
(ii) cloud shadows, which the cloud gate misses because it keys on bright pixels whereas a shadow is dark;
(iii) seamless intra-tile temporal overlap, a tile in which two regions are each individually clean and have no nodata line, so the automatic nodata and edge gates pass both halves, yet they come from acquisitions at different times, so illumination, sun angle and cast shadows disagree across the tile; no per-pixel or per-region threshold can flag this, since each part is valid on its own and only their juxtaposition is wrong;
(iv) dense fire smoke that occludes the surface like a cloud (the cloud gate does not flag it, being keyed on the Sen2Cor scene-classification cloud classes rather than on smoke), while thin or partial smoke is kept, since it is a legitimate part of the
scene; and 
(v) sensor-instrument artefacts that are not ground signal --- the bright cross-shaped SAR returns over large metallic structures, and the Sentinel-2 inter-band offset that splits a fast-moving object into displaced R/G/B copies --- removed only when pronounced enough to corrupt the tile. The visible nodata seam between two passes is the detectable counterpart of (iii) and is already removed by the nodata and edge gates of Phase~D; the manual step exists precisely for the seamless overlap, the shadows, the residual haze, the dense smoke and the sensor-instrument artefacts, which leave no per-pixel threshold signature. Each manual removal is recorded in $\Sigma$ under a single \texttt{manual\_review} category; the criterion that triggered it is not separately logged, but the recorded source identifiers let the decision be re-checked by re-fetching the tile. Flagged pairs are deleted, the corresponding stratum deficit is re-opened, and the offending source scene is marked as excluded so it will not be reused. Recovery then fills the re-opened deficits by searching, for each rejected/flagged location of an under-filled stratum, an alternative low-cloud optical scene in the same season (excluding scenes already marked as failed), pairing it with a SAR scene within $\pm\tau$, and re-applying the Phase~D quality gates. Every rejected attempt is eligible, whatever stopped it. We deliberately do not pre-classify failures into recoverable and non-recoverable: the measured outcome (Table~\ref{tab:rejections}) shows that the reason a tile was rejected is a poor predictor of whether a later attempt on the same ground will succeed, so the loop re-opens all of them and lets the retry decide.
The loop is bounded per location rather than by passes over the corpus: each reopened location is retried at most twice, and stops earlier when its candidate pool is exhausted --- no scene remains in that season that clears the quality gate and pairs within $\pm\tau$. The bound of two retries is set by the pairing window, not chosen arbitrarily: with a $\pm\tau$ window of three days, each attempt probes roughly a week of candidate dates, so two retries already sweep most of a three-month season; a third would begin to reach into the neighbouring season, where a match would carry the wrong stratum label, for little expected gain. Locations are visited in stratum-deficit order, so the effort concentrates where the shortfall is largest rather than being spread uniformly over the corpus.
The loop terminates when no pair is flagged and all strata have met their target, bounded by a maximum number of review rounds. The combination of (i) review-driven exclusion of bad sources and (ii) location-targeted recovery via alternative scenes makes this phase the quality and yield stage of DDSA: it removes what the automatic gates cannot see and refills what they rejected, while the stratum balance itself is held by the deficit-driven ordering that every removal feeds back into (Section~\ref{sec:formulation}).

Figure~\ref{fig:phaseE-examples} shows one representative removal per failure mode.

\subsection{Phase F --- Finalisation}
The accepted pairs are archived as the full corpus $D$. The reduced benchmark split $\dlite$ is derived by down-sampling every non-empty stratum to a common per-stratum cap, with strata below the cap entering whole; the exact cap and the sampling rule are recorded in the per-stratum statistics report $\Sigma$, which also exposes the per-stratum fill, rejection and recovery counts needed to characterise the corpus in Section~\ref{sec:datarecords}.

The complete procedure is given as pseudocode below: Algorithm~\ref{alg:ddsa1} covers the planning stage (Phases~A--C) and Algorithm~\ref{alg:ddsa2} the constrained acquisition with deficit recovery (Phases~D--F).

Table~\ref{tab:params} lists the settings used to build \corpus, so the symbols carry their realised values through the pseudocode.

\begin{table}[htbp]
	\centering
	\small
	\setlength{\tabcolsep}{4pt}
	\caption{Parameter settings used to build \corpus. Symbols match the phase descriptions above and Algorithms~\ref{alg:ddsa1}--\ref{alg:ddsa2}.}
	\label{tab:params}
	\begin{tabular}{>{\raggedright\arraybackslash}p{0.22\linewidth} >{\raggedright\arraybackslash}p{0.36\linewidth} >{\raggedright\arraybackslash}p{0.36\linewidth}}
		\toprule
		Setting & Value & Role \\
		\midrule
		Land-cover classes  & 8 of 11 (WorldCover v200)         & stratification axis \\
		Continental regions & 6                                 & stratification axis \\
		Seasons             & 4 (hemisphere-corrected)          & stratification axis \\
		Season window       & 3-month meteorological            & winter $Y$ = Dec\,$Y{-}1$--Feb\,$Y$ \\
		Pairs per season    & 2 (balanced)                      & atomic S1--S2 pairs per location \\
		Tile size           & $512\times512$ px @ 10\,m         & released tile \\
		Block size          & $\leq$2048 px ($\leq$50\,MB/req.) & acquire-once unit \\
		Target countries    & config-defined set               & geographic frame, batched \\
		$\tau$              & 3 days                            & S1--S2 temporal pairing window \\
		$c_m$              & 10\%                              & scene-metadata cloud pre-filter \\
		$c_r$              & 0.05                              & regional Sen2Cor SCL cloud \\
		$c_t$              & 0.03                              & per-tile bright-pixel ratio \\
		Edge-padding        & 0.15                              & max edge-artefact fraction \\
		Coverage / nodata   & $\geq$0.20 / $\leq$0.80 cand.; 0\% released & footprint gate / released purity \\
		$r$                & 0.30                              & max inter-block tile overlap \\
		$\lambda$          & $65^\circ$ (optical)              & optical limit binds the pair \\
		Seed                & 42                                & deterministic ordering \\
		\bottomrule
	\end{tabular}
\end{table}

\begin{algorithm}[t]
	\footnotesize
	\setlength{\algomargin}{1em}
	\DontPrintSemicolon
	\SetKwInOut{Input}{Input}
	\SetKwInOut{Output}{Output}
	\SetKwFunction{PhysRegion}{PhysicalRegion}
	\SetKwFunction{AggBlocks}{AggregateIntoBlocks}
	\SetKwFunction{ExtractMax}{ExtractMax}
	\SetKw{Continue}{continue}
	\SetKw{KwAnd}{and}
	
	\caption{DDSA --- Part 1: sampling frame, stratification and greedy maximum-coverage block plan}
	\label{alg:ddsa1}
	
	\Input{land-cover map $L$ (ESA WorldCover 2021, product v200; 10\,m);
		target per class $T$; seasons $S$, regions $R$;\\
		\hspace*{1em}per-sensor latitude limits $\lambda$
		(optical $\leq 65^{\circ}$, SAR $\leq 90^{\circ}$)}
	\Output{frame $F$; quota $q$; ordered $\mathit{plan}$;
		$\mathit{deficit}[\cdot]$}
	\BlankLine
	
	\phasebar{Phase A --- Sampling-frame construction}
	$F \leftarrow \emptyset$\;
	\ForEach{cell $t$ in the tiling of $L$ over the target countries}{
		$c_t \leftarrow \arg\max_{k}\,|\{\text{pixels of class }k\text{ in
		}t\}|$\tcp*{dominant class}
		$g_t \leftarrow \PhysRegion(\mathrm{lat}_t,\mathrm{lon}_t)$
		\tcp*{point-in-polygon}
		\lIf{$g_t \neq \textsc{OCEAN}$ \KwAnd $|\mathrm{lat}_t| \leq \lambda$}{$F \leftarrow F \cup \{(t,c_t,g_t,\mathrm{lat}_t,\mathrm{lon}_t)\}$}
	}
	\BlankLine
	
	\phasebar{Phase B --- Stratification and target quota}
	$\mathit{Strata} \leftarrow C \times S \times R$
	\tcp*{$C$: classes present in $F$}
	$q \leftarrow \lfloor T / (|R|\cdot|S|) \rfloor$\;
	\lForEach{stratum $s \in \mathit{Strata}$}{$\mathit{deficit}[s] \leftarrow q$}
	\BlankLine
	
	\phasebar{Phase C --- Greedy maximum-coverage block plan}
	$B \leftarrow \AggBlocks(F)$ \tcp*{acquire once, extract all tiles}
	\ForEach{block $b \in B$}{
		\ForEach{stratum $s$}{
			$\mathit{contrib}[b,s] \leftarrow |\{\text{tiles of }b\text{ in }s\text{
				over all seasons}\}|$\;
		}
	}
	$H \leftarrow$ max-heap of $B$ keyed by
	$\Pi(b)=\sum_{s}\min(\mathit{contrib}[b,s],\mathit{deficit}[s])$\;
	$\mathit{plan} \leftarrow \langle\,\rangle$\;
	\While{$H \neq \emptyset$ \KwAnd $\sum_{s}\mathit{deficit}[s] > 0$}{
		$b \leftarrow \ExtractMax(H)$\;
		$\Pi(b) \leftarrow
		\sum_{s}\min(\mathit{contrib}[b,s],\mathit{deficit}[s])$
		\tcp*{rescore: stale key}
		\lIf{$\Pi(b) \leq 0$}{\Continue}\tcp*{no contribution}
		append $b$ to $\mathit{plan}$\;
		\lForEach{stratum $s$}{$\mathit{deficit}[s] \leftarrow \max(0,\mathit{deficit}[s]-\mathit{contrib}[b,s])$}
	}
	\Return{$F,\,q,\,\mathit{plan},\,\mathit{deficit}$}\;
	
\end{algorithm}
\begin{algorithm}[t]
	\footnotesize
	\DontPrintSemicolon
	\SetKwInOut{Input}{Input}
	\SetKwInOut{Output}{Output}
	\SetKw{Continue}{continue}
	\SetKw{KwOr}{or}
	\SetKw{KwAnd}{and}
	
	\caption{DDSA --- Part 2: constrained acquisition and human-in-the-loop deficit recovery}
	\label{alg:ddsa2}
	
	\Input{$\mathit{plan}$, quota $q$ (from Algorithm~\ref{alg:ddsa1});\\
		\hspace*{1em}temporal tolerance $\tau$; cloud thresholds $c_m,c_r,c_t$;\\
		\hspace*{1em}max inter-block tile overlap $r$}
	\Output{full co-registered SAR--optical corpus $D$;\\
		\hspace*{1em}optional reduced benchmark split $\dlite$;\\
		\hspace*{1em}per-stratum statistics report $\Sigma$}
	\BlankLine
	
	\phasebar{Phase D --- Constrained block-level acquisition}
	\ForEach{block $b \in \mathit{plan}$, in dynamic stratum-priority order}{
		$Q \leftarrow \{\text{optical scenes over }b\text{ in season}:
		\text{cloud}\%\leq c_m\}$\;
		$o \leftarrow$ most-recent scene in $Q$, regional-cloud refined,
		excluding failed dates\;
		$a \leftarrow$ SAR scene with $|t(a)-t(o)|\leq\tau$
		\tcp*{temporal pairing}
		\lIf{$a=\emptyset$}{\Continue}
		download pair $(o,a)$\;
		\ForEach{tile $p$ extracted from $b$}{
			\uIf{$\mathrm{regionalCloud}(p)>c_r$ \KwOr
				$\mathrm{tileCloud}(p)>c_t$}{ reject $p$\; }
			\uElseIf{$p$ violates nodata/edge constraints}{ reject $p$\; }
			\uElseIf{$\exists$ accepted $q$ from another block in same season:
				$\mathrm{centroidDist}(p,q) < (1-r)\cdot\mathrm{footprint}$}{
				reject $p$\; }
			\Else{
				accept $p$;\quad $\mathit{counter}[s_p] \leftarrow
				\mathit{counter}[s_p]+1$\;
			}
		}
	}
	\Return{accepted pairs and counters; continued below}\;
\end{algorithm}
\addtocounter{algocf}{-1}
\begin{algorithm}[t]
	\footnotesize
	\DontPrintSemicolon
	\SetKwInOut{Output}{Output}
	\SetKw{Continue}{continue}
	\SetKw{KwOr}{or}
	\SetKw{KwAnd}{and}
	
	\caption{DDSA --- Part 2 (continued): human-in-the-loop deficit
		recovery and finalisation (Phases~E--F)}
	\label{alg:ddsa2b}
	
	\Output{full co-registered SAR--optical corpus $D$;
		reduced split $\dlite$; per-stratum report $\Sigma$}
	\BlankLine
	
	\phasebar{Phase E --- Human-in-the-loop QC with deficit repair (iterated)}
	\tcc{acquisition is coarse-to-fine: block-level bulk (D) maximises
		tiles+metadata per GEE call;}
	\tcc{tile-level recovery (E) maximises salvageable yield;}
	\tcc{each removal re-opens its stratum deficit, which the ordering follows}
	\Repeat{review flags no pair \KwAnd $\forall s:\mathit{counter}[s]\geq q$}{
		$R \leftarrow$ manual review of pairs accepted since the last round\;
		\ForEach{pair $p \in R$ flagged as residual-cloud, shadow, smoke,
			temporal-overlap or sensor artefact}{
			delete $p$;\quad
			$\mathit{counter}[s_p]\leftarrow\mathit{counter}[s_p]-1$\;
			$\mathit{deficit}[s_p]\leftarrow\mathit{deficit}[s_p]+1$\;
			mark the source scene of $p$ as excluded\;
		}
		\ForEach{stratum $s$ with $\mathit{counter}[s]<q$}{
			\tcp{recovery operates at tile/location granularity}
			\ForEach{rejected/flagged location of $s$, in stratum-priority order}{
				$o' \leftarrow$ alternative low-cloud optical scene, same season,
				excluding scenes already marked as failed\;
				$a' \leftarrow$ SAR scene with $|t(a')-t(o')|\leq\tau$\;
				\If{$(o',a')$ pass the Phase\,D quality gates}{
					accept the pair;\quad
					$\mathit{counter}[s]\leftarrow\mathit{counter}[s]+1$\;
					$\mathit{deficit}[s]\leftarrow\mathit{deficit}[s]-1$\;
				}
			}
		}
	}
	\tcp{bounded by a maximum number of review rounds}
	\BlankLine
	
	\phasebar{Phase F --- Finalisation}
	$D \leftarrow$ archive of accepted $(\text{optical},\text{SAR})$ pairs\;
	$\dlite \leftarrow$ down-sample $D$ to a common per-stratum cap
	\tcp*{strata below the cap enter whole}
	$\Sigma \leftarrow$ per-stratum fill, rejection and recovery statistics\;
	\Return{$D,\,\dlite,\,\Sigma$}\;
	
\end{algorithm}

%% file: sections/03_implementation.tex
\section{System Implementation}
\label{sec:implementation}

\begin{figure}[ht]
	\centering
	\resizebox{\linewidth}{!}{
	\begin{tikzpicture}[
		font=\sffamily,
		>={Stealth[round]},
		ext/.style   ={rounded corners=2pt, draw=green!45!black, fill=green!8,
			text=green!22!black, align=center, inner sep=4pt,
			text width=30mm, minimum height=10mm},
		stage/.style ={rounded corners=3pt, draw=black!55, fill=white,
			align=center, inner sep=4pt, text width=50mm,
			minimum height=11mm},
		state/.style ={rounded corners=2pt, draw=orange!60!black, fill=orange!10,
			text=orange!32!black, align=center, inner sep=4pt,
			text width=38mm, minimum height=10mm},
		corpus/.style={rounded corners=2pt, draw=orange!60!black, fill=orange!14,
			text=orange!28!black, align=center, inner sep=4pt,
			text width=58mm, minimum height=11mm},
		flow/.style  ={->, thick, black!65},
		link/.style  ={<->, dashed, black!55},
		feed/.style  ={->, black!45}
		]
		
		\node[stage] (s1)
		{\textbf{Pre-computation}\\[1pt]{\scriptsize tile indexing $\cdot$ block planning $\cdot$ scene precataloging}};
		\node[stage, below=9mm of s1] (s2)
		{\textbf{Acquisition}\\[1pt]{\scriptsize parallel GEE workers $\cdot$ rate-limit coordination\\ tile-level quality gates}};
		\node[stage, below=9mm of s2] (s3)
		{\textbf{Human-in-the-loop}\\[1pt]{\scriptsize review $\cdot$ decisions $\cdot$ tile-level recovery}};
		\node[stage, below=9mm of s3] (s4)
		{\textbf{Archival}\\[1pt]{\scriptsize hash-verified (xxHash64 + SHA256)\\ STAC / PDS4 / FAIR JSON sidecar}};
		
		\node[ext, above=14mm of s1.north, xshift=-2mm] (wc)
		{\textbf{ESA WorldCover}\\[1pt]{\scriptsize 10\,m land-cover labels}};
		\node[ext, left=4mm of wc] (ne)
		{\textbf{Natural Earth}\\[1pt]{\scriptsize admin-0 country polygons}};
		\node[ext, right=4mm of wc] (gee)
		{\textbf{Google Earth Engine}\\[1pt]{\scriptsize Sentinel-1 GRD $\cdot$ Sentinel-2 SR}};
		
		\node[state, right=24mm of s1] (db2)
		{\textbf{tile\_lookup.db}\\[1pt]{\scriptsize label-purity lookup}};
		\node[state, below=16mm of db2] (db1)
		{\textbf{pipeline.db}\\[1pt]{\scriptsize blocks $\cdot$ tiles $\cdot$ decisions $\cdot$ checkpoints}};
		
		\node[corpus, below=11mm of s4] (out)
		{\textbf{Released corpus}\\[1pt]{\scriptsize SAR--optical pairs + JSON sidecars\\ D (full) $\cdot$ D\_lite (reduced split)}};
		
		\begin{scope}[on background layer]
			\node[draw=green!45!black, dashed, rounded corners=3pt, fill=green!4,
			fit=(ne)(wc)(gee), inner sep=7pt] (extbox) {};
			\node[draw=black!45, rounded corners=3pt, fill=blue!3,
			fit=(s1)(s2)(s3)(s4), inner sep=8pt] (pipebox) {};
			\node[draw=orange!60!black, dashed, rounded corners=3pt, fill=orange!4,
			fit=(db1)(db2), inner sep=8pt] (statebox) {};
		\end{scope}
		
		\node[font=\sffamily\bfseries\scriptsize, text=green!25!black,
		anchor=north west] at (extbox.north west) {\,EXTERNAL DATA SOURCES};
		\node[font=\sffamily\bfseries\scriptsize, text=black!70,
		anchor=south west] at (pipebox.north west) {\,DDSA PIPELINE $\cdot$ Docker $\cdot$ idempotent \& resumable};
		\node[font=\sffamily\bfseries\scriptsize, text=orange!32!black,
		anchor=south west] at (statebox.north west) {\,STATE STORES $\cdot$ SQLite (WAL)};
		
		\draw[flow] (s1) -- (s2);
		\draw[flow] (s2) -- (s3);
		\draw[flow] (s3) -- (s4);
		\draw[flow] (s4) -- (out);
		
		\draw[feed] (ne.south)  to[out=-90,in=160] (s1.west);
		\draw[feed] (wc.south)  -- (s1.north);
		\draw[feed] (gee.south) to[out=-90,in=35]  (s1.east);
		
		\draw[link] (s1.east) -- (db2.west);
		\draw[link] (s2.east) -- (db1.west);
		\draw[link] (s3.east) to[out=20,in=-160] (db1.west);
	
	\end{tikzpicture}
	}
	\caption{DDSA system architecture. Three external data sources feed a Docker-containerised pipeline of four stages (pre-computation, acquisition, human-in-the-loop, archival), backed by two SQLite (WAL) state stores that drive scheduling and label-purity lookup. The released corpus is the bottom output.}
	\label{fig:system}
\end{figure}
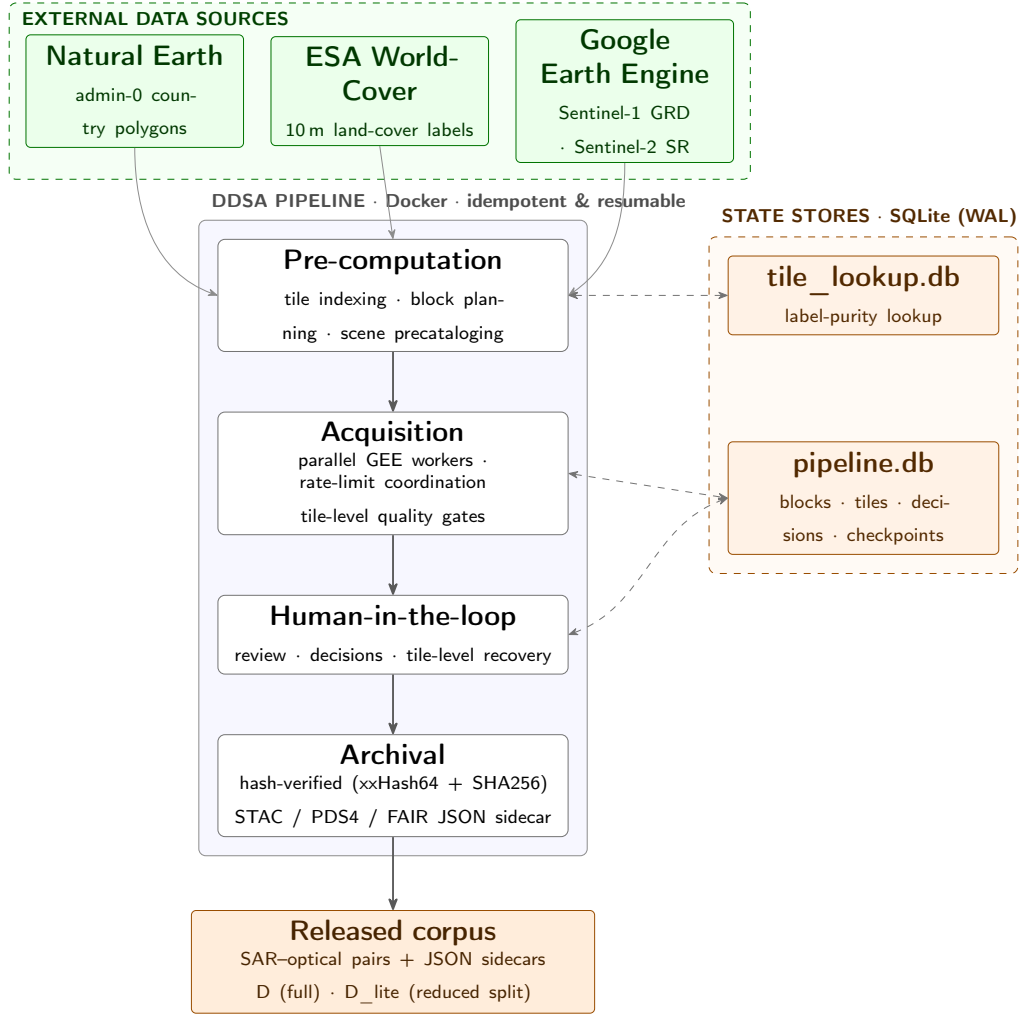

DDSA is realised as a Docker-containerised pipeline (Figure~\ref{fig:system}) designed for unattended execution on long, resource-bounded runs. Three properties drive the engineering decisions: idempotency, resumability, and provenance.

\paragraph{Containerisation and reproducibility}
The pipeline runs inside a versioned Docker image; the image tag is recorded in every per-pair sidecar, making it possible to retrieve the exact processing environment for any released pair. Dependencies are pinned.

\paragraph{State stores}
Two SQLite databases (operated in WAL mode for concurrent reader/writer safety) drive scheduling. \texttt{pipeline.db} holds the blocks, tiles, decisions and per-stage checkpoints; \texttt{tile\_lookup.db} holds the pre-computed per-tile label purity and composition, so each pair's sidecar can be populated without re-touching the WorldCover raster during acquisition. Stale processing flags are reset on resume so that interrupted runs do not leave orphans. Because acquisition is partitioned (see below), each batch keeps its own pair of databases; once a batch is closed they are consolidated, unchanged, into a single read-only master database from which every statistic reported in this paper is computed. The master keeps the delivered pairs, the pairs that were acquired and later withdrawn, and the attempts an automatic gate stopped, in three separate tables, so the product and the cost of producing it are never conflated in a count.

\paragraph{Parallel GEE access}
Acquisition runs as a pool of parallel workers against Google Earth Engine. A rate-limit coordinator throttles requests below GEE's per-account quotas; failed requests are retried with exponential backoff, and persistently failing scenes are added to a no-coverage cache so they are not re-attempted in later rounds or in Phase~E recovery. 

\paragraph{Per-country partitioning}
Because the sampling frame is delimited by target country, acquisition can be run one country at a time rather than over the whole frame at once. In practice the corpus was built in per-country batches, which keeps each run small, makes progress easy to checkpoint and resume, and confines any re-run to the country being repaired. The same partitioning also lends itself to distributing the acquisition across independent workers; in this work parallelism stayed within a single provider's per-account rate limits (see above), and any wider distribution would remain subject to the same quota constraints.

\paragraph{Archival}
Integrity is recorded at two levels: each pair carries a fast, non-cryptographic xxHash64 checksum in its sidecar (Section~\ref{sec:sidecar}) for quick post-release corruption checks by the user, and the release ships a SHA256 manifest over the distributed archives for cryptographic-grade verification. Each released pair carries its standards-aligned JSON sidecar.

During dataset assembly, this same integrity layer flagged \AlgoNCorruptPairs\ pairs whose files had been corrupted in storage or transfer --- a small fraction at this data volume --- which were removed from the release; the checksums shipped with the corpus let any user re-run the same verification.

\paragraph{Geographic and geodetic corrections}
Block bounding boxes are computed with a $\cos(\mathrm{lat})$ correction~\citep{snyder1987map}; the antimeridian is handled by splitting affected blocks. Hemisphere correction is applied at the stratification stage so that the season axis is meaningful in both hemispheres.
Co-registration is by construction rather than by post-hoc alignment: for each tile we request both modalities over the same footprint and write them to the same grid --- identical bounding box, CRS (the tile's local UTM zone) and 10\,m pixel size --- so the optical and SAR rasters of a pair share pixels one to one. The two scenes are matched within a temporal window of at most $\tau$ days (here three); the realised gap is stored per pair (\texttt{temporal\_gap\_days}, alongside the two acquisition timestamps), so the pairing can be audited directly from the sidecar.

\paragraph{Post-acquisition integrity audit}
After Phase~E terminates, a fixed-set integrity audit re-verifies the released corpus against six spatial-and-temporal invariants: spatial uniqueness (no pixel-identical duplicates at the same position), cross-class boundary consistency, temporal-source uniqueness within a season (within a season, no location is sampled twice from the same source scene --- a cross-tile check, not a guarantee that a single tile draws from one acquisition only), monthly sampling diversity, the inter-block tile overlap
constraint of Phase~D, and intra-block extraction-window non-collision. Any violation is logged and the offending pair is removed atomically (disk first, database last); a machine-readable audit report is shipped with the released corpus. The audit checks quantities that are measurable per pixel or across tiles; a defect such as a seamless within-tile temporal overlap, whose two halves are each individually valid and leave no per-pixel signature, falls outside its reach and is handled instead by the Phase~E human review (Section~\ref{sec:methods}). The audit is the defence-in-depth counterpart to the acquisition-time guards.

%% file: sections/04_corpus.tex
\section{The \corpus Corpus}
\label{sec:datarecords}

The release, \corpus, consists of the full corpus $D$, the derived reduced benchmark split $\dlite$, the per-stratum statistics report $\Sigma$, the post-acquisition integrity audit report, the pre-trained SSL backbone weights of Section~\ref{sec:validation}, and the full method specification given by this paper and the pseudocode of Algorithms~\ref{alg:ddsa1} and~\ref{alg:ddsa2}.

\subsection{Folder hierarchy}
Pairs are stored as a parsing-free access layer: the directory path encodes the stratum, so most filtering is a file-system operation and needs no metadata parsing. The tree is split first by sensor and then by stratum, 
\begin{quote}\small
	\texttt{<sensor>/class\_<class>/season\_<season>/region\_<region>/}
\end{quote}
for example \texttt{s2/class\_built\_up/season\_summer/region\_AFRICA/}. Picking a class, season or region is then a directory walk (\texttt{cp -r s2/class\_built\_up/}\,\dots), and restricting to a single modality is just choosing the \texttt{s1} or \texttt{s2} root. The two members of a pair sit at the matching \texttt{s1/} and \texttt{s2/} paths and share the same \texttt{tile\_id} and \texttt{pair\_id}, so optical--SAR matching is a lookup on those identifiers --- which are also carried in the filename and the sidecar. The layout trades a little redundancy on disk for a large drop in friction: common selections need no loader, no index and no metadata I/O.

\subsection{Per-pair sidecar}
\label{sec:sidecar}
Every released tile is accompanied by a per-tile JSON sidecar that records the full geographic, dimensional, classification, source-data, acquisition, integrity, license and provenance context of the tile. The schema is fixed: it is aligned with STAC~1.0.0~\citep{stac2021spec}, informed by the NASA PDS4 information model~\citep{nasapds4}, and designed to satisfy the FAIR principles~\citep{wilkinson2016fair}; the top-level fields \texttt{stac\_compliant} and \texttt{stac\_version} record the targeted STAC version and the alignment state. Top-level identifiers \texttt{tile\_id}, \texttt{pair\_id}, \texttt{filename}, and \texttt{sensor} link the tile to its geographic location, to its modality counterpart in the pair, and to its on-disk filename. The schema is organised in nine blocks:

\begin{itemize}
	\item \textbf{integrity}: \texttt{checksum} (xxHash64 of the .tif),
	\texttt{algorithm}, and a descriptive note. Designed for fast, non-cryptographic corruption detection on the released file.
	\item \textbf{coordinates}: centroid \texttt{lat},\texttt{lon} (WGS84 datum), geographic \texttt{bbox}, and \texttt{crs} -- the native projected CRS (e.g.\ \texttt{EPSG:32754} for UTM zone 54S over Australia). Both WGS84 and the local projection are recorded so downstream tooling can choose either reference frame.
	\item \textbf{dimensions}: \texttt{width} and \texttt{height} in pixels, \texttt{pixel\_resolution\_m} (10), the list of source \texttt{bands} (e.g.\ B4/B3/B2 for a Sentinel-2 true-colour tile; VV, VH and a precomputed VV$-$VH difference for Sentinel-1), \texttt{dtype}, and the \texttt{nodata} sentinel value. For Sentinel-2 the field records \texttt{uint16}, the native type of the source product; the released optical files hold the same integer digital numbers stored as \texttt{float32}, which is also the type of the Sentinel-1 files.
	\item \textbf{classification}: the hemisphere-corrected season triple (\texttt{season}, \texttt{hemisphere}, \texttt{season\_raw}), the dominant land-cover class (\texttt{land\_cover\_class}, integer \texttt{class\_id}), the
	\texttt{geographic\_region} assigned via Natural Earth, the \texttt{label\_purity} (ratio of the dominant class), the full \texttt{label\_composition} as a list of $(\texttt{class\_id}, \texttt{class\_name}, \texttt{ratio})$ entries, and \texttt{label\_source}
	(\texttt{ESA\_WorldCover\_2021\_v200}). Since a season may span two calendar years, the \texttt{season} field is the authoritative stratum key; the \texttt{satellite\_acquisition\_time} recorded in the \textbf{acquisition} block gives the date and should not be used on its own to infer the season.
	\item \textbf{data\_source}: ingestion \texttt{provider} (\texttt{Google Earth Engine}), source \texttt{mission} (\texttt{ESA Copernicus Programme}), GEE \texttt{collection} identifier (\texttt{COPERNICUS/S2\_SR} or \texttt{COPERNICUS/S1\_GRD}), and \texttt{processing\_level} (L2A for Sentinel-2, GRD for Sentinel-1).
	\item \textbf{acquisition}: \texttt{download\_timestamp} (when DDSA downloaded the tile, UTC ISO~8601), \texttt{satellite\_acquisition\_time} (when the satellite captured the scene), specific \texttt{spacecraft} (e.g.\ Sentinel-2B),
	\texttt{max\_cloud\_percentage} (the scene-level $c_m$ applied), and \texttt{temporal\_pairing\_}\allowbreak\texttt{tolerance\_days}~$=\tau$.
	\item \textbf{license}: \texttt{terms} (CC BY 4.0), \texttt{source} (\texttt{Copernicus Sentinel data}), attribution \texttt{url} and \texttt{citation}.
	\item \textbf{provenance}: the exact GEE asset identifiers \texttt{sentinel2\_image\_id} and \texttt{sentinel1\_image\_id} of the source scenes, the producing \texttt{script} and \texttt{script\_version}, the \texttt{dataset\_version}, the \texttt{processing\_timestamp}, the \texttt{docker\_image}, and the \texttt{processing\_node} identifier. Together these allow any tile to be traced back to its source scenes and to a
	re-instantiable processing environment.
	\item \textbf{quality\_metrics}: \texttt{temporal\_gap\_days} (the realised $|t(a)-t(o)|$ for the SAR--optical pair).
\end{itemize}

A complete example sidecar is shipped in the released \texttt{examples/} folder so downstream tooling can be validated against a concrete reference.

\subsection{Filename encoding}
The filename of each released tile encodes the same stratum information present in the folder hierarchy and the sidecar, so that downstream tooling can identify a tile's class, region, season, sensor, pairing and download timestamp from a directory listing alone -- no metadata I/O required. The format is:

 \begin{center}
 	\small\texttt{<sensor>\_<class>\_<season>\_<region>\_<tile\_id>\_<timestamp>\_<pair\_id>.tif}
 \end{center}

\noindent For example, the filename

\begin{quote}\small
	\texttt{S2\_CROP\_SUM\_AUS\_00cb7d7d92d3\_20260411T174932\_2ac26529158baed0.tif}
\end{quote}

\noindent decomposes as:
\begin{itemize}[leftmargin=2em]
	\item sensor \texttt{S2}; class \texttt{CROP} (cropland);
	season \texttt{SUM} (summer); region \texttt{AUS} (Australia/Oceania);
	\item \texttt{tile\_id} \texttt{00cb7d7d92d3}, a 12-character identifier;
	\item download timestamp \texttt{20260411T174932};
	\item \texttt{pair\_id} \texttt{2ac26529158baed0}, a 16-character identifier.
\end{itemize}

\medskip

The \texttt{pair\_id} is shared between the Sentinel-2 file and the matched Sentinel-1 file of the same pair, so optical--SAR matching reduces to a string match on the filename. Its width is set at 64 bits so that accidental repetition stays negligible at corpus scale: a shorter identifier would be easier to read but would make coincidental repeats likely once the corpus passes a few hundred thousand pairs, at which point tooling keying on \texttt{pair\_id} alone could silently mismatch two unrelated files. The released corpus bears this out: its {\AlgoNPairs} identifiers are pairwise distinct, with no collision to resolve. The \texttt{tile\_id} is carried in the filename beside it, so a pair is in any case addressable by the composite $(\texttt{tile\_id}, \texttt{pair\_id})$.
Should two pairs ever derive the same identifier, the consolidation step keeps both under a suffixed \texttt{pair\_id} (\texttt{\dots\_1}) propagated to the four files and to the two sidecars, rather than letting one displace the other. No such collision occurs in this release --- all identifiers are distinct --- but a reader parsing filenames should treat the final field as an identifier that may carry a suffix.

The timestamp is the download-batch time; the exact satellite-acquisition time and the per-file write time are recorded in the sidecar (Section~\ref{sec:sidecar}).

\subsection{Band selection and extensibility}
Each released tile carries the three Sentinel-2 true-colour bands (B4/B3/B2) and three Sentinel-1 channels --- the polarisations VV and VH and a precomputed $\mathrm{VV}-\mathrm{VH}$ difference. Because the difference is stored rather than computed on the fly, the six-channel input used in Section~\ref{sec:validation} is read straight from the released files. This is a deliberate choice, not a loss of information. Shipping the full Sentinel-2 spectral stack and the complete Sentinel-1 product would multiply both the on-disk size and, more importantly, the pre-processing and pre-training cost of the corpus to the point where only a few very well-resourced groups could actually use it. Since the contribution of this work is the acquisition algorithm, and a corpus is only useful if people can train on it, we release the bands that are the most widely used and the most directly interpretable:
human-readable true colour for the optical modality and the standard VV/VH polarisations for SAR. The radiometry of both modalities follows the Google Earth Engine collections they were drawn from. Sentinel-2 values are the Level-2A surface-reflectance digital numbers of \texttt{COPERNICUS/S2\_SR}, which, unlike its harmonised successor, keeps the radiometric offset that ESA introduced with processing baseline 04.00 in January 2022; since every acquisition in the corpus is later than that change, all optical tiles carry it, and surface reflectance is recovered as $(\mathrm{DN}-1000)/10^{4}$. Sentinel-1 values are GEE's GRD backscatter in dB, after thermal-noise removal, radiometric calibration and terrain correction, with no speckle filtering; only interferometric-wide, dual-polarisation (VV+VH) scenes are admitted, from either orbit direction, and the third channel is the VV$-$VH difference in dB, the logarithm of the co- to cross-polarised ratio.

The choice does not lock users out of the remaining bands. Every sidecar records the exact source-scene identifiers (\texttt{sentinel2\_image\_id}, \texttt{sentinel1\_image\_id}) together with the footprint and the acquisition time, so any additional Sentinel-1 or Sentinel-2 band can be recovered per tile from the same Google Earth Engine assets with a small, well-defined amount of work. The released corpus is therefore a practical default that runs on commodity hardware while remaining fully extensible to the complete spectral and polarimetric stack; producing that full-band version over the worked regions is left as future work.

The same metadata that makes the missing bands recoverable also makes the corpus easy to reshape for a given study. Because every tile records its native CRS, footprint, pixel resolution and stratum, a user can filter the release down to the classes, regions or seasons they care about, re-tile the imagery into smaller patches, or pool it to a coarser resolution (for example 20\,m or 50\,m) without going back to the source archives. This is the point of the release: \corpus is a worked product that shows what the algorithm produces and gives others something concrete to build on, but the contribution of the paper is the acquisition algorithm itself, not any single fixed form of its output.

\subsection{Label-source caveat}
\label{sec:labelcaveat}
The following must be stated explicitly in any use of the dataset and is recorded in the released README and Dataset Card: 
\begin{itemize}
	\item Labels are derived from ESA WorldCover 2021, product version v200 	(\texttt{ESA\_WorldCover\_2021\_v200}; \citealp{zanaga2022worldcover}).
	\item A temporal gap of about four years between the WorldCover product and the Sentinel acquisitions may introduce label noise.
	\item Acquisitions belong to the 2025 season cycle; because the winter window opens in December of the preceding year, part of the corpus carries December~2024 dates. The \texttt{season} field, not the calendar year, is the stratum key.
	\item Labels are intended for self-supervised learning where they serve as stratification only, not as ground truth for downstream supervised tasks.
\end{itemize}

The stratification labels are derived from ESA WorldCover 2021 (v200), the most recent validated global land-cover product at 10\,m resolution and built from the same Sentinel-1 / Sentinel-2 sources as this corpus. All Sentinel acquisitions used here belong to the 2025 season cycle. Because the winter window opens in December of the preceding year (Section~\ref{sec:methods}), part of the corpus carries December~2024 acquisition dates; this affects \AlgoDecPairs\ of \AlgoNPairs\ released pairs (\AlgoDecPct\%) --- December is winter in the Northern hemisphere and summer in
the Southern, so both season labels appear among them. The gap between the label product and the imagery is therefore about four years throughout, and users who group the corpus by calendar year rather than by the \texttt{season} field will split that winter in two. This gap does not affect the validity of the pairs or their intended use. In this corpus the land-cover class is not a supervised target: it is used only as a stratification key, to stratify acquisition across the (class $\times$ season $\times$ region) strata. Users who need ground-truth labels for a supervised task are expected to bring their own from external benchmarks, as the released Dataset Card notes. Its residual effect is limited to the small fraction of tiles whose dominant class has changed since 2021 --- almost entirely in areas of real land-cover change --- and it touches only the exact stratum assignment of those tiles, not the quality of the pair or its usefulness for self-supervised pre-training. We use WorldCover rather than a more recent annual land-cover product because it is independently validated, widely adopted as a reference, and produced from the same sensors; for a stratification key we consider this more important than exact temporal alignment.

\subsection{Quality-gate rejection taxonomy}
\label{sec:rejections}
Rejections are reported in $\Sigma$ under a fixed taxonomy in two layers. The acquisition cascade (Table~\ref{tab:rejections}) logs one row per pair attempt, under the reason that stopped it: tile-level quality rejections --- \texttt{clouds} (bright-pixel fraction above $c_t$), \texttt{nodata} (missing pixels), \texttt{edge\_padding} (scene-edge artefacts); image-wide rejections, where the source scene itself is unusable --- \texttt{has\_inf} (non-finite pixels), \texttt{striped\_corruption} (truncated download), \texttt{flat} (single-value image), \texttt{orphan\_missing\_file} (recorded tile whose file is missing); and \texttt{pool\_exhausted}, recorded when no candidate scene remained for a location under the required constraints.

\begin{figure}[htbp]
	\centering
	\includegraphics[width=\linewidth]{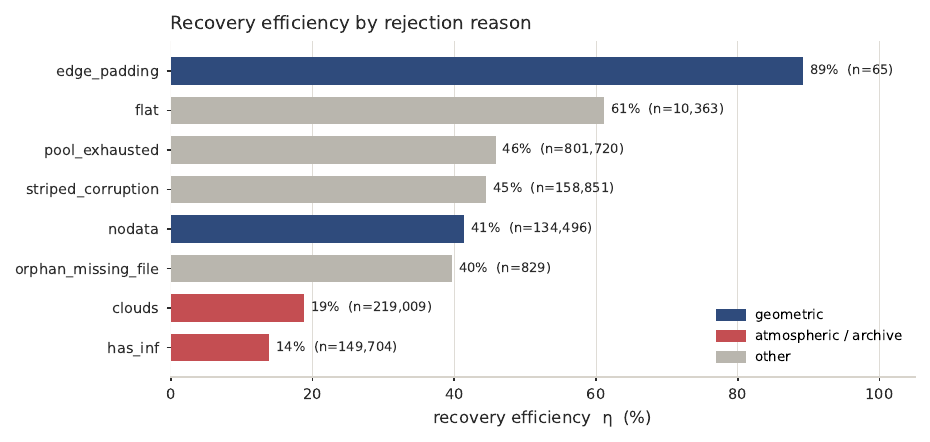}
	\caption{Recovery efficiency $\eta$ by rejection reason, ordered by $\eta$. Bars are coloured to contrast the two families the text compares, geometric (\texttt{nodata}, \texttt{edge\_padding}) and atmospheric/archive (\texttt{clouds}, \texttt{has\_inf}), with the remaining reasons in grey, and annotated with the number of retried attempts behind each ratio. The ordering inverts the intuitive reading of the taxonomy: \texttt{nodata}, which looks terminal, is recovered more than twice as often as \texttt{clouds}, which sits near the bottom --- a pattern consistent with recoverability following the temporal persistence of the cause rather than the kind of defect. Thin categories, whose $\eta$ rests on few retries, are read against their annotated count rather than taken as stable estimates.}
	\label{fig:reason-eta}
\end{figure}

Because the repair stage updates the rejection row it retried rather than writing a new one, each row also carries the outcome of that retry, and Table~\ref{tab:rejections} reports it beside the count. The result contradicts the intuitive reading of the taxonomy and is worth stating plainly: \emph{the reason an attempt was rejected is a poor predictor of whether it can be recovered.} A geometric failure, which looks terminal, is recovered more than twice as often as a cloud rejection --- \texttt{nodata} at $\eta={\AlgoEtaNodataPct\%}$ --- plausibly because a Sentinel-1 pass on a different orbit covers a different footprint, so a missing corner need not be missing on the next attempt. The atmospheric and archive failures, which look transient, are the least recoverable --- \texttt{clouds} at $\eta={\AlgoEtaCloudsPct\%}$ and \texttt{has\_inf} at ${\AlgoEtaHasInfPct\%}$ --- consistent with a location that is cloudy being cloudy on most dates of its season, a persistence that global cloud climatologies make explicit~\citep{wilson2016cloud}. The reading that fits both is that what separates them is not the kind of defect but whether its cause persists in time: an orbital geometry changes between passes, a climate does not. If it holds, it is also the mechanism behind the geographic dropout the algorithm exists to counteract. We read this as observational evidence consistent with temporal persistence rather than as a proven causal mechanism: the categories differ in $\eta$ and also, unavoidably, in where their locations fall, so part of the spread could reflect that geometric failures happen to sit in clearer places. What we can say from scene metadata alone is that the reason a gate stopped an attempt is a poor predictor of whether a retry succeeds; the persistence reading is the most parsimonious explanation, not a controlled result (Figure~\ref{fig:reason-eta}).
The \texttt{edge\_padding} efficiency rests on {\AlgoNRejEdge} rejections of
which only {\AlgoNRetriedEdge} were ever retried, and is reported for completeness rather than as a stable estimate.
After acceptance, two further reasons apply at tile level: removals raised by the post-acquisition audit, reported under the invariant they breach --- \texttt{overlap\_geometric} (two tiles of the same season sharing more than $r=0.30$ of their footprint), \texttt{season\_quota} (a location carrying more pairs in one season than the configured limit) and \texttt{month\_repeat} (two pairs of the same location and season drawn from the same calendar month, which the interleaved month
order is meant to prevent) --- and a single \texttt{manual\_review} category for the Phase~E human removals. The reviewer applies the criteria listed in Phase~E, but the specific criterion per tile is not separately logged, as hand-labelling a small fraction by sub-type is disproportionate; the recorded source identifiers let any removed tile be re-fetched and the decision re-checked. Reporting at this granularity lets the user recompose subsets along quality dimensions without re-running the
acquisition.

\begin{table}[htbp]
	\centering
	\small
	\setlength{\tabcolsep}{4pt}
	\caption{Acquisition rejection counts by reason, from the \texttt{rejected\_tiles} log. Counts are per acquisition attempt, not per unique tile: a location may fail several times before a valid pair is found, so the total exceeds the number of released tiles. The left block is the rejection; the right block is what the repair loop then did with it --- how many of those attempts it refilled, how many it retried without success, and the resulting efficiency $\eta$ (Equation~\eqref{eq:recovery}). Reading the two together is the point of the table: $\eta$ varies more than fourfold across reasons, and not in the direction the reason suggests. Pairs removed after acceptance by the manual review do not appear here; they are logged as withdrawn pairs rather than as gate rejections and are counted in the scale-and-effort summary. Attempts not yet retried are excluded from $\eta$ and reported there as well. Percentages are over the total number of \emph{rejected} attempts, so the table is closed over its own rows and sums to 100 up to rounding; the wider count of every attempt the acquisition made (delivered, withdrawn and rejected) is reported separately in the scale-and-effort summary.} 
	\label{tab:rejections}
	\input{tables/rejections_table}
\end{table}

\paragraph{Recovery yield}
The unit is the pair attempt, which is the atomic unit of the acquisition: one row of the rejection log is one attempt at one pair, and the repair loop records its outcome on that same row. Writing $N_{\mathrm{rec}}$ for the attempts it refilled and $N_{\mathrm{exh}}$ for those it retried without success, the recovery efficiency is
\begin{equation}
	\label{eq:recovery}
	\eta = \frac{N_{\mathrm{rec}}}{N_{\mathrm{rec}} + N_{\mathrm{exh}}},
\end{equation}
over all {\AlgoNRejected} rejected attempts. Attempts not yet retried are excluded from both terms and reported apart, so that an operational limit --- quota, time --- does not present itself as a failure of the method. Realised values:  $N_{\mathrm{rec}}={\AlgoNEtaRec}$ and $N_{\mathrm{exh}}={\AlgoNEtaExh}$, giving $\eta={\AlgoEtaPct\%}$, with {\AlgoNEtaPending} attempts not yet retried. The per-reason $\eta$ of Table~\ref{tab:rejections} is read against its own retry count, reported in the same row: where only a few attempts of a category were ever retried --- \texttt{edge\_padding} is the extreme case, at {\AlgoNRetriedEdge} retries over {\AlgoNRejEdge} rejections --- the ratio is indicative rather than a stable estimate, and we say so in the table rather than averaging the thin categories into the headline figure. 

Counted in pairs rather than attempts, the share of $D$ that exists only because of the loop is ${\AlgoNRecovered}/{\AlgoNPairs}={\AlgoRepairSharePct\%}$ --- the pair-level counterpart of the ``no repair'' column of Table~\ref{tab:coverage} --- and \AlgoQCSurvivalPct\%\ of those pairs survived the subsequent manual review, so the repair does not buy coverage at the cost of quality. Attempts and pairs are different units and are not netted against one another: a single delivered location can absorb several refilled attempts, so the {\AlgoNLedgerSuccess} refilled attempts do not equal the {\AlgoNRecovered} recovered pairs they leave in $D$. The released corpus excludes both the pairs the manual review later withdrew and those the cross-batch consolidation removed as redundant coverage.

\begin{figure}[htbp]
	\centering
	\includegraphics[width=\linewidth]{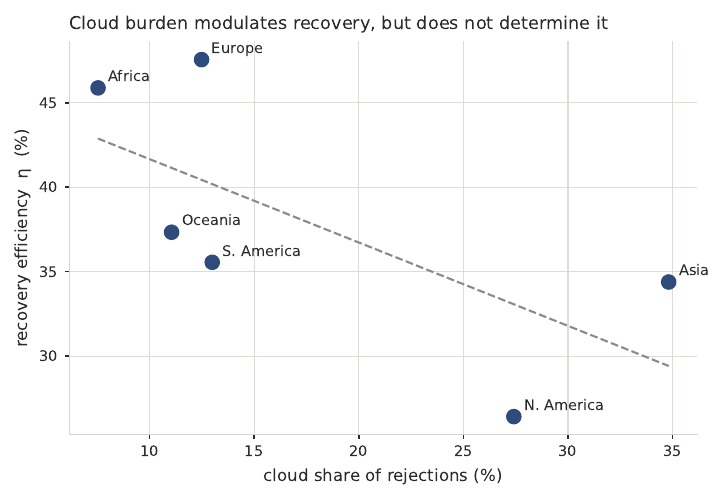}
	\caption{Per-region recovery efficiency $\eta$ against cloud burden, the share of a region's rejections that were cloud-driven. Cloud burden modulates recovery but does not determine it: the cloudier regions tend lower, yet the relation is not monotone --- North America breaks it, with a moderate cloud share but the lowest efficiency --- so what bounds recovery is regional scene availability and orbital geometry rather than cloud alone. The dashed line is a descriptive trend, not a fit.} 
	\label{fig:recovery-cloud}
\end{figure}

Table~\ref{tab:recovery} breaks $\eta$ down by region and by latitude band, and the two cuts tell different stories that must not be conflated. Across the tropical boundary the efficiency is essentially flat --- it is marginally higher inside the tropics --- while the exhausted attempts concentrate in the same band: the loop does not perform worse where cloud persists, it runs out of usable scenes. Across regions, by contrast, the efficiency varies widely, from \AlgoEtaNorthAmericaPct\% in North America to \AlgoEtaEuropePct\% in Europe. That spread is wider than the latitudinal one, so what moves $\eta$ is regional scene availability rather than latitude as such. Cloud burden modulates it but does not determine it: Figure~\ref{fig:recovery-cloud} plots each region's efficiency against the share of its rejections that were cloud-driven, and while the cloudier regions do tend lower, the relation is not monotone --- North America has only a moderate cloud share yet the lowest efficiency of all --- which points to availability and orbital geometry, not cloud alone, as the binding factor. Two readings are ruled out by the same table. The retry effort was not spread uniformly: deficit-ordered visiting sent proportionally more retries to the tropical band and left a larger share of extratropical attempts untried, so the flat latitudinal efficiency holds despite an uneven allocation of effort rather than because the allocation was balanced. And the attempts whose footprint could not be resolved to a latitude are fewer than one in a thousand, so the band split is complete in practice. 

\begin{table}[htbp]
	\centering
	\small
	\setlength{\tabcolsep}{4pt}
	\caption{Phase-E repair by scope. The left block counts pair attempts and yields $\eta$ (Equation~\eqref{eq:recovery}); the right block counts delivered pairs and the share of each scope's corpus that the loop supplied. The two blocks are in different units and are not additive.}
	\label{tab:recovery}
	\input{tables/recovery_table}
\end{table}

\subsection{Scale and effort summary}
The figures below are computed from the consolidated master database; the rejection counts of Table~\ref{tab:rejections} come from the same source. 
\begin{itemize}
	\item Released pairs: {\AlgoNPairs}, over {\AlgoNTiles} distinct locations;
	\item Stratification: 8 land-cover classes $\times$ 4 hemisphere-corrected seasons $\times$ 6 continental regions, sampled over a configurable set of target countries (bounded into small download batches); all {\AlgoStratK} strata are populated, with a mean of {\AlgoStratMean} pairs per stratum and {\AlgoStratAboveHundred} of them above one hundred. The full range is {\AlgoStratMin} to {\AlgoStratMax}: the low tail is the signature of the stopping point rather than a gap in coverage. Because the fill order is driven by the per-stratum deficit, every stratum receives samples before any one saturates, so a longer acquisition grows the sparse strata first and the range narrows monotonically with compute time. Per-stratum counts are in $\Sigma$;
	\item Total acquisition attempts: {\AlgoNAttempts} in all (delivered pairs, pairs withdrawn after acceptance, and attempts stopped by an automatic gate); the rejected share is broken down by category in Table~\ref{tab:rejections};
	\item Pairs supplied by the Phase~E repair loop: {\AlgoNRecovered} ({\AlgoRepairSharePct\%} of $D$); attempts retried: {\AlgoNEtaRec} refilled and {\AlgoNEtaExh} exhausted, with {\AlgoNEtaPending} not yet retried. The loop is bounded per location (at most two retries, or fewer if the candidate pool runs out) rather than by a number of passes over the corpus; 
	\item Removed after acceptance: {\AlgoNWithdrawn} pairs in all, which re-enter the deficit like any other failure. Of these, {\AlgoNManual} pairs over {\AlgoNManualTiles} distinct locations were removed during acquisition, \num{80734} by human review and the remainder by an automated post-download integrity check that discards orphaned, truncated or otherwise corrupted downloads; the other \num{13} were withdrawn by the post-acquisition integrity audit. The audit removals are itemised in $\Sigma$, together with the cross-batch consolidation, which is not counted here;
	\item On-disk size of the full corpus: \num{2.64}~TB, held as {\AlgoNRasterFiles} raster files (one optical and one radar per pair), each with a JSON sidecar beside it. The reduced split $\dlite$ holds \num{20916} pairs over all {\AlgoStratK} strata, capped at 112 per stratum; strata holding fewer pairs than the cap contribute all they have, so the split is capped rather than strictly equalised. On disk it occupies \num{76.5}~GB. 
	
\end{itemize}	

The target countries are a configurable set rather than a fixed list: instead of treating the whole planet as one frame, the acquisition is bounded to small batches of countries, which keeps each download manageable and lets the worked area grow batch by batch across the six inhabited continents. The exact set is recorded in the released configuration and Dataset Card.

\subsection{Coverage comparison against opportunistic corpora}
\label{sec:coverage-comparison}

Table~\ref{tab:features} situates \corpus among these corpora. At the level of content and scale the corpora are comparable; what sets \corpus apart is its acquisition, which samples an explicit (class, season, region) stratification with a per-stratum quota enforced during the run and re-prioritises the strata that cloud cover depletes, where the reference corpora sample an opportunistically collected pool and filter it afterwards.

\input{tables/feature_comparison_table}

DDSA is meant to cover every (class~$\times$~season~$\times$~region) stratum by
construction, rather than the imbalance that opportunistic collection tends
to leave in multi-modal SAR--optical corpora. The question is therefore not
whether the corpus spans more of the planet than a global archive, but
whether deficit-driven acquisition fills the strata more evenly than an
opportunistic download would over the same frame. To check that this
actually happens, we compare the coverage of the released corpus against two corpora that
were assembled opportunistically, SSL4EO-S12~\citep{wang2023ssl4eo} and SEN12MS~\citep{schmitt2019sen12ms}. Each of these corpora was built for a purpose other than balanced coverage: SSL4EO-S12 is an unlabelled pre-training corpus sampled around the most populated cities, so its frame follows population rather than land cover, and its authors note that requiring cloud-free scenes biases where its locations end up~\citep{wang2023ssl4eo}, whereas SEN12MS samples land and urban areas at random yet, as its own authors report, still under-represents some land-cover classes such as water~\citep{schmitt2019sen12ms}. Everything here is computed from public scene metadata --- centroid coordinates and acquisition season --- and from the per-stratum report $\Sigma$ of Section~\ref{sec:methods}; nothing is trained and no extra imagery is downloaded.

The thing that matters most is that we re-stratify the two reference corpora under our own scheme, not theirs. For each scene we assign a land-cover class by point lookup on the same ESA WorldCover 2021 product~\citep{zanaga2022worldcover}, a continental region by point-in-polygon intersection with Natural Earth admin-0 boundaries~\citep{naturalearth}, and a hemisphere-corrected season from the  acquisition month. The strata are therefore identical across all three corpora, and the contrast is not an artefact of different labelling conventions.

\begin{figure}[htbp]
	\centering
	\includegraphics[width=\linewidth]{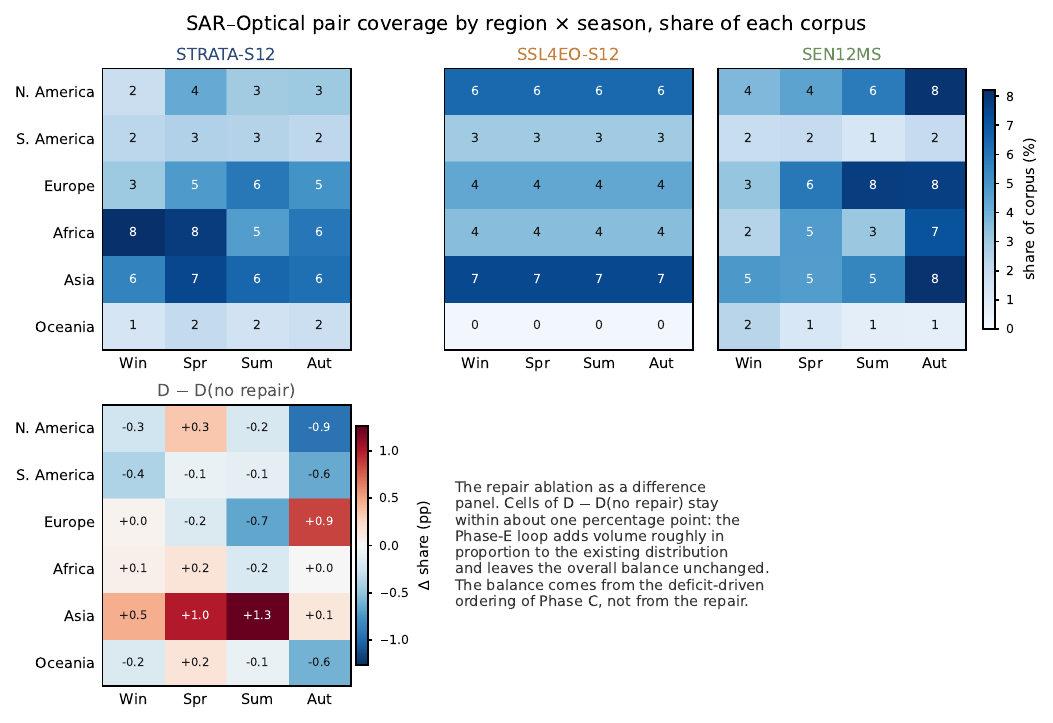}
	\caption{Coverage of \corpus against SSL4EO-S12 and SEN12MS over the shared (class~$\times$~season~$\times$~region) strata. Each top panel is one corpus; cell colour is the share of that corpus's samples falling in a (region, season) stratum, so the panels are comparable across corpora that differ in size by orders of magnitude, and the seasonal axis is read left-to-right within a panel. The lower-left panel is the $D - D(\text{no repair})$ difference: its cells stay within about one percentage point, showing that the Phase-E repair loop adds volume roughly in proportion to the existing distribution rather than redistributing it, so the stratum balance comes from the deficit-driven ordering of Phase~C upstream (cf.\ Table~\ref{tab:coverage} and Section~\ref{sec:coverage-comparison}).}
	\label{fig:coverage}
\end{figure}

\begin{figure}[htbp]
	\centering
	\includegraphics[width=\linewidth]{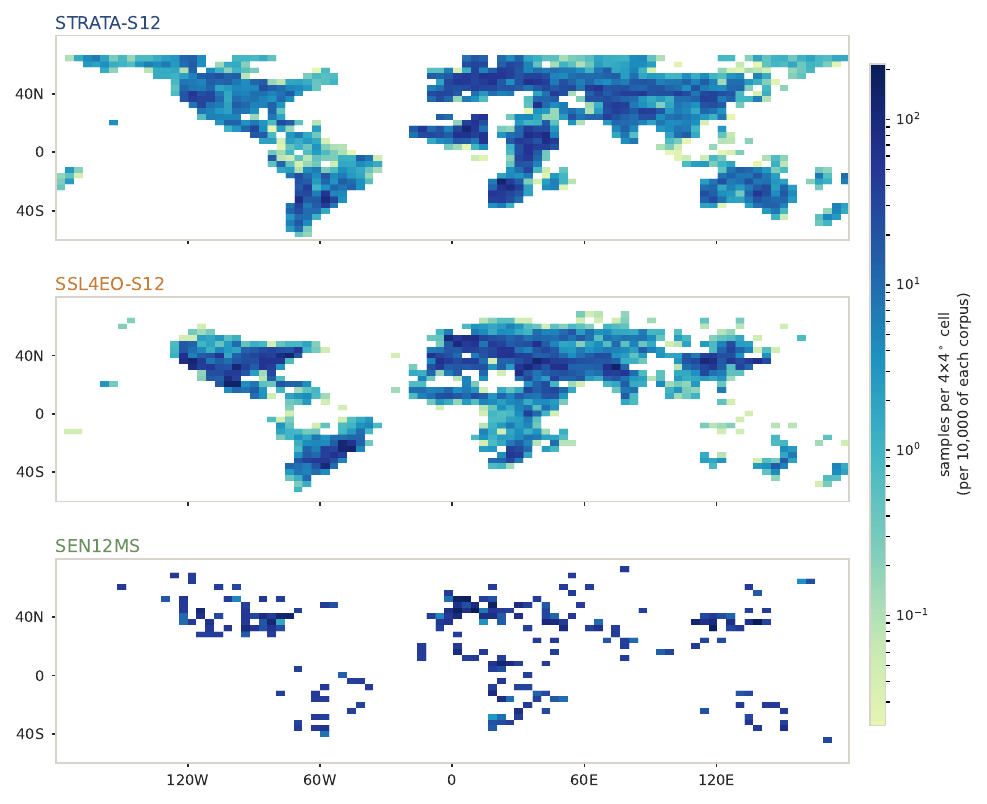}
	\caption{Global sample density of the three corpora, binned on a $4^\circ$ grid and normalised to each corpus so the panels compare distribution rather than size. The colour scale is logarithmic and shared: densities span several orders of magnitude between an empty cell and a dense continental one, and on a linear scale everything but the hottest cells collapses into the background. The three regimes are visibly different: SSL4EO-S12 concentrates its densest cells along the northern mid-latitudes and thins below the equator, the signature of a frame that follows population; SEN12MS is sparse everywhere, being two orders of magnitude smaller; \corpus spreads across the worked continents with its densest cells in equatorial Africa, the Amazon and southern Asia. The empty land areas of the top panel are countries this acquisition has not reached yet, not strata it failed to fill (Section~\ref{sec:limitations}).}
	\label{fig:map}
\end{figure}

\begin{figure}[htbp]
	\centering
	\includegraphics[width=\linewidth]{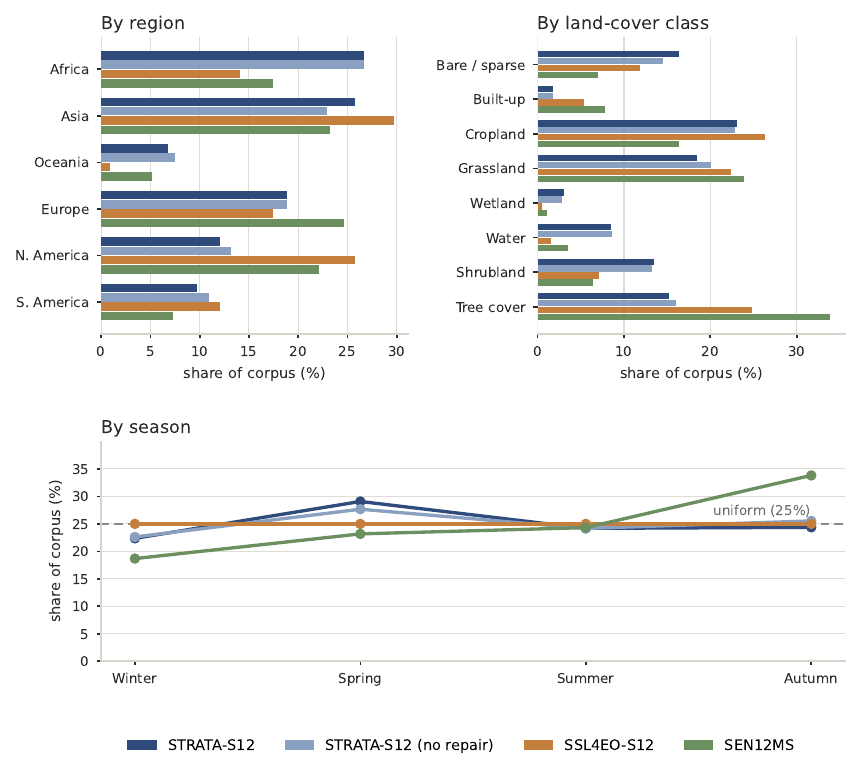}
	\caption{Marginal share of each corpus over the three stratification axes. Bars are percentages of each corpus's own samples, so corpora differing in size by more than an order of magnitude remain comparable. The three marginals are shown in place of the full $K={\AlgoStratK}$ grid, which is unreadable at this width and answers the same question less directly: which continent, which land cover and which season each acquisition regime over- or under-represents. The season panel is drawn as a profile because the axis is an ordered cycle, with the uniform share marked so departures are read against a reference rather than against a truncated axis.}
	\label{fig:profiles}
\end{figure}

We report four descriptors over the strata: (i) the Northern-to-Southern hemisphere sample ratio; 
(ii) the number of land-cover classes that reach at least 1\% of a corpus's samples, counted over the eight WorldCover classes that the shared stratification retains (snow and ice, mangroves, and moss and lichen do not occur across all sampled regions and are dropped in Phase~B); 
(iii) the normalised Shannon entropy of the stratum-occupancy distribution, which is higher when samples are spread across many strata instead of piling up in a few; and (iv) the fraction of samples at tropical latitudes ($|\mathrm{lat}| \le 23.5^{\circ}$), the band that cloud cover hits hardest. Table~\ref{tab:coverage} collects the result, and Figure~\ref{fig:coverage} shows the region$\times$season share of each corpus as a heatmap, with the repair ablation drawn as a difference panel. Figures~\ref{fig:map} and~\ref{fig:profiles} then open the four descriptors up: the first shows where each corpus actually sampled, the second how each distributes its samples over region, land cover and season. The table summarises; these two are where the summary comes from.

One caveat applies to all of them. Every sample counted here for \corpus has passed the Phase-D gate and the Phase-E review, at a cost of {\AlgoNRejected} rejected attempts for {\AlgoNPairs} released pairs; each reference corpus was filtered under its own criteria, from a scene-level cloud threshold in SSL4EO-S12 to cloud-scored mosaicking followed by expert inspection of every scene and patch in SEN12MS~\citep{wang2023ssl4eo,schmitt2019sen12ms}. The comparison is therefore of how each regime distributes what it kept, not of equally conditioned samples.

The two figures separate what the four descriptors compress. Geographically (Figure~\ref{fig:map}) the difference is not a matter of degree: SSL4EO-S12 places its densest cells along the northern mid-latitudes and thins sharply south of the equator, while \corpus carries substantial mass across equatorial Africa, the Amazon basin and southern Asia. The regional marginals (Table~\ref{tab:regionshare}; Figure~\ref{fig:profiles}) put numbers on it --- Africa holds \num{26.7}\% of \corpus against \num{14.1}\% of SSL4EO-S12, and Oceania \num{6.8}\% against \num{0.9}\% --- with the complementary deficit in North America, which the population-following frame  favours and the deficit-driven one does not.

\begin{table}[htbp]
	\centering
	\small
	\setlength{\tabcolsep}{6pt}
	\caption{Regional share of each corpus (\% of that corpus's own samples) over the six continental regions of the shared frame. Columns sum to 100 up to rounding.}
	\label{tab:regionshare}
	\input{tables/region_share_table}
\end{table}

Across land-cover classes the contrast is one of spread rather than of any single class. \corpus ranges from \num{2}\% to \num{23}\% of its samples across the eight classes, against \num{1}\% to \num{33.5}\% for SEN12MS, which is the entropy difference of Table~\ref{tab:coverage} read directly.
Two classes are worth naming. Permanent water bodies take \num{8.5}\% of \corpus against \num{1.7}\% and \num{3.5}\%, which quantifies the water under-representation the SEN12MS authors report of their own corpus~\citep{schmitt2019sen12ms}. Built-up is the one class \corpus is sparsest on (\num{2}\%), since the per-class target favours classes that dominate a tile and built-up dominates comparatively few tiles of the frame.
The seasonal axis is the one on which this corpus does not lead. SSL4EO-S12 acquires four seasonal snapshots per location and is therefore uniform across seasons by construction, whereas \corpus carries a mild spring excess and SEN12MS a pronounced autumn one. Season is a stratification axis here, not a per-location requirement, so the quota is met per stratum rather than per site; matching a per-location seasonal sampling would be a change of acquisition policy rather than of balance control, and is not claimed.

One descriptor is deliberately not among the four. Stratum occupancy --- the fraction of the $K={\AlgoStratK}$ cells holding at least one sample --- saturates at this scale and no longer separates the corpora: all three exceed 98\% and two reach 100\%. With hundreds of thousands of samples spread over {\AlgoStratK} cells, even a strongly skewed sampler touches every cell at least once, so occupancy measures corpus size more than acquisition strategy. We report it as supplementary and base the comparison on how the mass is distributed within the frame, which is what the entropy and the geographic fractions capture. The comparison is also read with the acquisitions' different stopping points in mind: the geographic fractions are shares and are insensitive to where a run was stopped, whereas entropy and occupancy are not.

\begin{table}[htbp]
	\centering
	\small
	\setlength{\tabcolsep}{4pt}
	\caption{Coverage distribution of the DDSA corpus against opportunistic corpora, computed from public scene metadata over the shared (class~$\times$~season~$\times$~region) stratification. Higher normalised entropy and more classes covered indicate a broader spread across strata. The hemisphere ratio is read with care: land area is itself Northern-skewed, so part of any imbalance is geographic, but DDSA's ratio is driven by its per-stratum quotas rather than by scene availability, so a strong skew in an opportunistic corpus points to cloud-induced dropout beyond that baseline.}
	\label{tab:coverage}
	\input{tables/coverage_table}
\end{table}

This is a descriptive comparison, and we are careful not to oversell it: the corpora were built over different sampling frames and target regions, so the contrast reflects how each was acquired rather than one cleanly controlled variable. To isolate the algorithm on its own we add an internal ablation over our own sampling frame. The column ``DDSA $D$ (no repair)'' recomputes the same descriptors over only the Phase-D bulk acquisition, leaving out the tiles that the Phase-E deficit-repair loop brought back. With the frame held fixed, the gap between the two columns isolates what the repair contributed to the distribution.
The ablation is a filter over the released corpus rather than a second acquisition: every delivered pair records whether the Phase-E repair loop brought it back, so the ``no repair'' column is the same corpus with those pairs withheld, scored by the same code on the same frame. Nothing is re-downloaded and no parameter changes between the two columns.

The result is not the one the framing would predict, and we report it as measured: withholding the repaired pairs leaves the four descriptors essentially unchanged, and marginally more favourable. The reading is consistent with Table~\ref{tab:rejections}. The loop succeeds where an alternative clean scene exists, and an alternative clean scene exists where the sky is clear --- which is where the bulk acquisition had already succeeded. It therefore adds volume roughly in proportion to the existing distribution rather than redistributing it. Read together with Equation~\eqref{eq:recovery}, the two quantities answer different questions and should not be conflated: $\eta$ measures how much of what was retried was refilled, while the ablation measures how much that refilling changed the shape of the corpus. A loop can score well on the first and leave the second flat, and here it does. What the repair buys, on this evidence, is {\AlgoRepairSharePct\%} more pairs at unchanged distribution and unchanged quality --- yield rather than balance --- while the balance itself comes from the deficit-driven ordering of Phase~C, upstream of the repair.

\subsection{Isolating the acquisition ordering}
\label{sec:phasec-ordering}

The coverage comparison of Section~\ref{sec:coverage-comparison} contrasts \corpus with corpora built over different frames, so it cannot, on its own, separate what the deficit-driven ordering of Phase~C contributes from what the choice of frame contributes. We isolate the ordering with a controlled replay over a single frame and a single attempt record: every acquisition attempt the run made, \SimTotalAttempts\ in total, successes and failures alike. Each attempt keeps its stratum and its real outcome --- a delivered pair, a gate rejection, or a pair withdrawn after acceptance --- so a policy that keeps returning to a cloudy stratum pays for the failures there, and the budget is counted in attempts rather than in delivered pairs. Per-stratum success rates in this record range from \SimPSuccessMin\% to \SimPSuccessMax\%. Within each stratum the outcomes are shuffled and consumed without replacement, so a policy can only spend attempts that really happened. We compare three orderings: \emph{deficit-greedy}, the Phase~C rule, which attempts next in the stratum with the fewest delivered pairs; \emph{round-robin}, which gives every stratum one attempt in turn regardless of what it has delivered, i.e.\ uniform effort without deficit bookkeeping; and \emph{random}, which draws attempts uniformly from the pooled record. Each is replayed over ten seeds. At each budget we report the number of the $K={\AlgoStratK}$ strata holding at least one pair, the normalised occupancy entropy, and the pairs delivered, together with the budget at which all {\AlgoStratK} strata first hold a pair.

\begin{figure}[t]
	\centering
	\includegraphics[width=\linewidth]{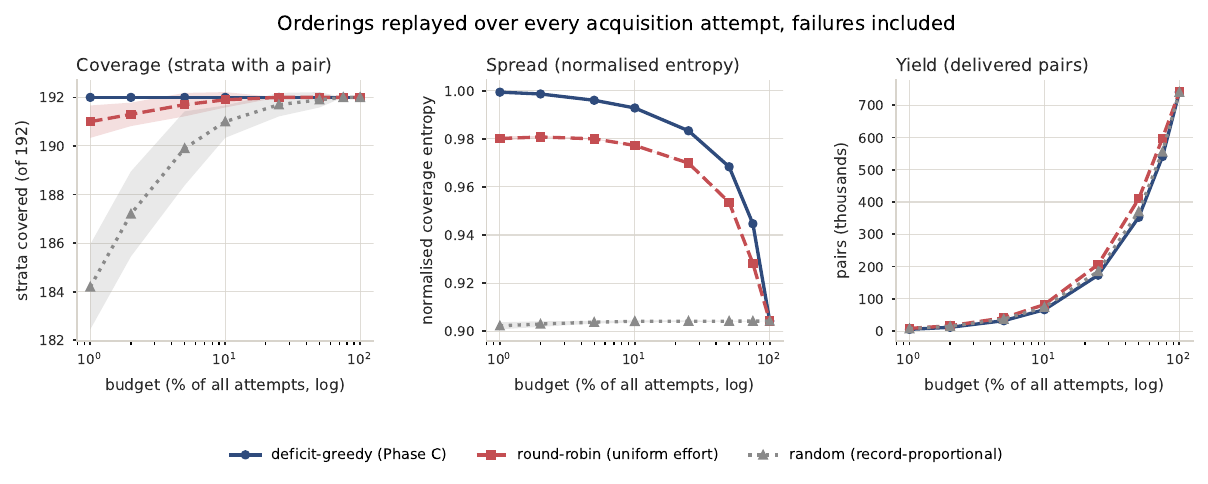}
	\caption{Coverage versus budget for the three orderings replayed over every acquisition attempt, failures included (mean over ten seeds; shaded band, one standard deviation). Left: strata holding at least one pair. Centre: normalised coverage entropy. Right: delivered pairs. Deficit-greedy covers every stratum from the start and stays the most even, at the cost of a lower yield at partial budgets; the three converge at the full budget because they spend the same record (Section~\ref{sec:phasec-ordering}).}
	\label{fig:phasec}
\end{figure}

Table~\ref{tab:phasec} and Figure~\ref{fig:phasec} report the result. Deficit-greedy populates every stratum after \SimFullDeficit\% of the attempts (about \SimFullDeficitAttempts\ attempts; \SimFullDeficitMin--\SimFullDeficitMax\% over seeds), against \SimFullRoundRobin\% for round-robin (\SimFullRoundRobinMin--\SimFullRoundRobinMax\%) and \SimFullRandom\% for random (\SimFullRandomMin--\SimFullRandomMax\%). It is also the most even at every partial budget: at 5\% of the attempts its normalised entropy is \SimDeficitEntropyFive, against \SimRoundRobinEntropyFive\ for round-robin and \SimRandomEntropyFive\ for random. Uniform effort is not enough on its own, because equal attempts become unequal pairs once success rates differ by two orders of magnitude; tracking the deficit is what redirects effort to the strata that convert it worst. All three orderings converge at the full budget, as they must, since they spend the same record. The reading is that the stratum balance of \corpus is a property of the acquisition \emph{ordering}, not of how far the run was taken: a partial deficit-driven run is already a covering one.

The balance has a price, and the replay measures it. Because it spends attempts where they convert worst, deficit-greedy delivers fewer pairs per attempt at a partial budget: at 5\% of the attempts, \SimDeficitPairsFive\ pairs against \SimRandomPairsFive\ for random (\SimYieldCostFive\% fewer) and \SimRoundRobinPairsFive\ for round-robin (\SimYieldCostRRFive\% fewer); at 25\% the gap to random narrows to \SimYieldCostTwentyFive\%. Two caveats bound the result. The replay works at stratum level and abstracts away blocks, one of which can feed several strata at once. And it can only spend attempts that DDSA chose to make; that record already over-samples the rare strata, so the random baseline inherits part of DDSA's own targeting. An acquisition driven by availability alone would have attempted the rare strata less often still, which makes the measured gap a conservative estimate. This finite-budget behaviour is consistent with the structure of the planning objective (Section~\ref{sec:formulation}): because the coverage function is monotone and submodular and each block's marginal gain is clipped to the residual deficit, greedy selection spreads early mass across the poorest strata rather than deepening the richest.

\begin{table}[htbp]
	\centering
	\small
	\setlength{\tabcolsep}{4pt}
	\caption{Acquisition orderings replayed over the full attempt record (\SimTotalAttempts\ attempts, failures included), over the shared $K={\AlgoStratK}$ strata; mean over ten seeds, $\pm$ one standard deviation where it is not negligible. Deficit-greedy is the Phase~C rule; round-robin gives each stratum one attempt in turn; random draws from the pooled record. All three spend the same attempts, so they coincide at the full budget. The bottom block gives the budget at which every stratum first holds a pair.}
	\label{tab:phasec}
	\input{algo_evidence/sim/sim_table}
\end{table}

\subsection{What the corpus demonstrates about the algorithm}
\label{sec:algorithm-evidence}

The released corpus is the algorithm's product, and the measurements of the preceding subsections are evidence about the algorithm rather than about this particular corpus. None of them depends on how far the acquisition was run. Two of the results have already been established above and are only recalled here: that a partial deficit-driven run is already a covering one, shown by the ordering experiment of Section~\ref{sec:phasec-ordering} (Table~\ref{tab:phasec}, Figure~\ref{fig:phasec}); and that the type of a defect is a poor predictor of its recoverability, with temporal persistence as the most parsimonious explanation, read from the per-reason efficiencies of Section~\ref{sec:rejections} (Table~\ref{tab:rejections}). What remains to state here is the third result, which the other subsections leave implicit: that the repair loop buys its \AlgoRepairSharePct\%\ of the corpus at no measurable cost in quality.

Three quantities, all computed over the delivered pairs, support that claim. First, the pairs the loop recovered survive the subsequent manual review at \AlgoQCSurvivalPct\%, a review in which the gallery did not show which pairs came from recovery, so recovery does not simply defer removal to a later stage. Second, their temporal pairing is no looser than that of the first-pass pairs: the mean SAR--optical gap is \AlgoGapRecovered\ days for recovered pairs against \AlgoGapFirstPass\ days for first-pass ones, well within the $\tau=3$ window either way. Third, their land-cover purity is not lower --- \AlgoPurityRecovered\ against \AlgoPurityFirstPass\ --- so the alternative scene the loop finds is, if anything, slightly better centred on its dominant class than a first-pass one. The repair therefore adds volume in the difficult strata without relaxing the guarantees that define a delivered pair; taken with the ordering result, this is why the loop is worth running even though, as Section~\ref{sec:coverage-comparison} showed, it does not by itself redistribute the corpus. That redistribution is the job of the deficit-driven ordering upstream; the loop's job is yield at constant quality.

The corpus itself is a worked instance of these statements, not the claim. It is offered as a usable product --- and we expect it to be used --- but its role in this paper is to make the results above measurable rather than asserted.

%% file: tables/rejections_table.tex
\begin{tabular}{lrrrrr}
\toprule
& \multicolumn{2}{c}{Rejected} & \multicolumn{3}{c}{Repair loop} \\
\cmidrule(lr){2-3}\cmidrule(lr){4-6}
Reason & $n$ & \% & recovered & exhausted & $\eta$ (\%) \\
\midrule
\multicolumn{6}{l}{\emph{Tile-level quality gates}} \\
\quad\texttt{clouds} & 444,136 & 21.4 & 41,131 & 177,878 & 18.8 \\
\quad\texttt{nodata} & 256,183 & 12.3 & 55,722 & 78,774 & 41.4 \\
\quad\texttt{edge\_padding} & 318 & 0.0 & 58 & 7 & 89.2 \\
\midrule
\multicolumn{6}{l}{\emph{Image-wide corruption (source scene unusable)}} \\
\quad\texttt{has\_inf} & 236,457 & 11.4 & 20,833 & 128,871 & 13.9 \\
\quad\texttt{striped\_corruption} & 313,707 & 15.1 & 70,724 & 88,127 & 44.5 \\
\quad\texttt{flat} & 16,574 & 0.8 & 6,338 & 4,025 & 61.2 \\
\quad\texttt{orphan\_missing\_file} & 1,776 & 0.1 & 329 & 500 & 39.7 \\
\midrule
\multicolumn{6}{l}{\emph{Availability (no candidate scene remained)}} \\
\quad\texttt{pool\_exhausted} & 809,884 & 39.0 & 367,612 & 434,108 & 45.9 \\
\midrule
Total rejected attempts & 2,079,035 & 100.0 & 562,747 & 912,290 & 38.2 \\
\bottomrule
\end{tabular}

%% file: tables/recovery_table.tex
\begin{tabular}{lrrrrr}
\toprule
& \multicolumn{3}{c}{Pair attempts} & \multicolumn{2}{c}{Delivered pairs} \\
\cmidrule(lr){2-4}\cmidrule(lr){5-6}
Scope & recovered & exhausted & $\eta$ (\%) & from repair & share (\%) \\
\midrule
Global & 562,747 & 912,290 & 38.2 & 292,432 & 39.5 \\
\quad Africa & 147,611 & 174,121 & 45.9 & 78,309 & 39.6 \\
\quad Asia & 158,733 & 302,889 & 34.4 & 88,156 & 46.2 \\
\quad Australia/Oceania & 29,001 & 48,679 & 37.3 & 16,863 & 33.3 \\
\quad Europe & 122,321 & 134,892 & 47.6 & 55,541 & 39.7 \\
\quad North America & 62,933 & 175,307 & 26.4 & 30,719 & 34.3 \\
\quad South America & 42,148 & 76,402 & 35.6 & 22,844 & 31.8 \\
\quad tropical band & 169,862 & 262,704 & 39.3 & 91,671 & 37.8 \\
\bottomrule
\end{tabular}

%% file: tables/feature_comparison_table.tex
\begin{table*}[htbp]
	\centering
	\scriptsize
	\caption{Feature comparison of \corpus against the most widely used multi-modal SAR--optical corpora for self-supervised EO, with the optical-only BigEarthNet as a reference point. Figures for the reference corpora are from their original publications. (a) content and scale; (b) acquisition design, where \corpus controls per-stratum coverage during the run while the reference corpora filter an opportunistically collected pool afterwards.}
	\label{tab:features}
	
	\renewcommand{\arraystretch}{1.25}
	\begin{tabularx}{\linewidth}{@{} >{\raggedright\arraybackslash}p{0.17\linewidth} >{\hsize=1.3\hsize\centering\arraybackslash}X *{3}{>{\hsize=0.9\hsize\centering\arraybackslash}X} @{}}
		\toprule
		\textbf{(a) Content and scale} & \textbf{\corpus} & \textbf{SSL4EO-S12} & \textbf{SEN12MS} & \textbf{BigEarthNet} \\
				& (this work) & \citealp{wang2023ssl4eo} & \citealp{schmitt2019sen12ms} & \citealp{sumbul2019bigearthnet} \\
		\midrule
		Modalities         & S1 + S2 & S1 + S2 & S1 + S2 + MODIS & S2 only \\
		Pairs / patches    & \num{740746} & ${\sim}\num{1}$M\textsuperscript{a} & \num{180662} & \num{590326} \\
		Distinct locations & \num{473786} & \num{251079} & 252 scenes\textsuperscript{c} & 125 tiles \\
		S2 bands           & 3\textsuperscript{b} & 13 & 13 & 12 \\
		S1 bands           & 3 & 2 & 2 & --- \\
		Resolution         & 10\,m & 10\,m & 10\,m & 10\,m \\
		Tile size          & $512^2$ & $264^2$ & $256^2$ & $120^2$ \\
		File layout        & stratum-encoded folders and filenames & GeoTIFF per band, by product and season & season / scene folders, patch ID in filename & one folder per patch, tile and date in name \\
		Licence            & CC BY 4.0 & CC BY & CC BY & CDLA-Perm. 1.0 \\
		\bottomrule
	\end{tabularx}
	
	\vspace{2pt}
	{\scriptsize
		\textsuperscript{a}~\num{251079} locations $\times$ 4 seasonal snapshots.\quad
		\textsuperscript{b}~True colour; full stack recoverable per tile (Section~\ref{sec:sidecar}).\quad
		\textsuperscript{c}~\num{180662} patches cut with 50\% overlap from 252 scenes.}
	
	\vspace{10pt}
	
	\renewcommand{\arraystretch}{1.15}
	 \begin{tabularx}{\linewidth}{@{} >{\raggedright\arraybackslash}p{0.17\linewidth} >{\hsize=1.3\hsize\raggedright\arraybackslash}X *{3}{>{\hsize=0.9\hsize\raggedright\arraybackslash}X} @{}}
		\toprule
		\textbf{(b) Acquisition design} & \textbf{\corpus} & \textbf{SSL4EO-S12} & \textbf{SEN12MS} & \textbf{BigEarthNet} \\
		\midrule
			Sampling frame    & class, season, region strata & populated cities & random over land and urban areas & 10 EU countries \\
			Coverage control  & per-stratum quota, online & per-city density & scene tiling & scene tiling \\
			Cloud dropout     & deficit re-opened, stratum re-prioritised & location re-drawn & cloud-free mosaics; dropped scenes not replaced & flagged, dropped \\
			Temporal pairing  & strict $\pm\tau$ gate & $\pm$30 days of four seasonal dates & same seasonal window & n/a (S2 only) \\
					Quality control   & cloud, nodata, edge and corrupt-scene gates + visual review of every pair & scene cloud cover $<$10\%; overlap filter & cloud-scored mosaicking + expert check of scenes and patches & tile cloud cover $<$1\%; snow, cloud and shadow patches flagged visually \\
			Per-pair metadata & FAIR STAC--PDS4 sidecar & per-patch metadata file & GeoTIFF georeference & per-patch labels and metadata JSON \\
		\bottomrule
	\end{tabularx}
\end{table*}

%% file: tables/region_share_table.tex
\begin{tabular}{@{}lrrr@{}}
\toprule
Region & STRATA-S12 & SSL4EO-S12 & SEN12MS \\
\midrule
N. America & 12.1 & 25.8 & 22.1 \\
S. America & 9.7 & 12.0 & 7.3 \\
Europe & 18.9 & 17.5 & 24.7 \\
Africa & 26.7 & 14.1 & 17.5 \\
Asia & 25.8 & 29.7 & 23.2 \\
Oceania & 6.8 & 0.9 & 5.2 \\
\bottomrule
\end{tabular}

%% file: tables/coverage_table.tex
\begin{tabular}{lcccc}
\toprule
Descriptor & DDSA $D$ & DDSA $D$ (no repair) & SSL4EO-S12 & SEN12MS \\
\midrule
Hemisphere ratio (N:S) & 2.92 & 2.74 & 4.99 & 4.09 \\
Land-cover classes covered & 8 & 8 & 7 & 8 \\
Normalised coverage entropy & 0.904 & 0.912 & 0.871 & 0.864 \\
Tropical-band fraction (\%) & 32.7 & 33.6 & 18.9 & 22.7 \\
\bottomrule
\end{tabular}

%% file: algo_evidence/sim/sim_table.tex
\begin{tabular}{l l r r r}
\toprule
Budget & Ordering & Strata covered & Norm.\ entropy & Pairs \\
\midrule
1\% & deficit-greedy (Phase~C) & 192 & 0.999 & \num{5199} \\
 & round-robin (uniform effort) & 191.0 $\pm$ 0.7 & 0.980 & \num{8198} \\
 & random (record-proportional) & 184.2 $\pm$ 1.8 & 0.902 & \num{7400} \\
\midrule
5\% & deficit-greedy (Phase~C) & 192 & 0.996 & \num{32056} \\
 & round-robin (uniform effort) & 191.7 $\pm$ 0.5 & 0.980 & \num{40960} \\
 & random (record-proportional) & 189.9 $\pm$ 1.5 & 0.904 & \num{37025} \\
\midrule
25\% & deficit-greedy (Phase~C) & 192 & 0.983 & \num{173229} \\
 & round-robin (uniform effort) & 192 & 0.970 & \num{206268} \\
 & random (record-proportional) & 191.7 $\pm$ 0.5 & 0.904 & \num{185042} \\
\midrule
100\% & deficit-greedy (Phase~C) & 192 & 0.904 & \num{740746} \\
 & round-robin (uniform effort) & 192 & 0.904 & \num{740746} \\
 & random (record-proportional) & 192 & 0.904 & \num{740746} \\
\midrule
\multicolumn{5}{l}{\emph{All 192 strata covered at (\% of attempts; mean [min--max])}} \\
 & deficit-greedy (Phase~C) & \multicolumn{3}{r}{0.05 [0.03--0.10]} \\
 & round-robin (uniform effort) & \multicolumn{3}{r}{4.03 [0.45--11.51]} \\
 & random (record-proportional) & \multicolumn{3}{r}{20.53 [6.86--52.88]} \\
\bottomrule
\end{tabular}

%% file: sections/05_validation.tex
\section{Technical Validation}
\label{sec:validation}

The aim of this section is deliberately narrow: to show that the released corpus is usable for what we built it for, self-supervised pre-training, and nothing beyond that. We do not claim that \corpus outperforms other corpora, nor that the method we use is the best available; those questions belong to the follow-up work mentioned earlier. We also stop short of one neighbouring claim: we do not test here whether the broader stratum coverage reported in Section~\ref{sec:coverage-comparison} improves downstream performance, since that benefit-of-coverage question requires a controlled cross-corpus comparison with geographically held-out evaluation, which we leave to the follow-up work. What we report is a single, pre-registered check that the corpus carries enough signal to learn from. The probe is an in-domain transfer test --- reBEN is optical/SAR land-cover imagery, like the corpus --- which is the standard evaluation regime for SAR--optical SSL~\citep{fuller2023croma,wang2024decur} and the same one SSL4EO-S12 uses~\citep{wang2023ssl4eo}. We take that as the appropriate setting rather than a limitation: the two arms differ only in whether the encoder saw $D$, so a random-init backbone put through the identical in-domain probe is the control that isolates what pre-training on the corpus contributed. That the pre-trained arm clears the criterion while the random one does not is therefore attributable to the corpus, not to the ease of the task.

\subsection{Pre-registered protocol}
We fix the protocol before running anything and record it as a dated, read-only artefact: the metric, the version of the downstream dataset, the splits, the random seeds, the pre-training schedule, and the numerical criterion for success. Only one quantity is set by experiment rather than in advance --- the number of pre-training epochs, which we calibrate once on the target hardware from a short timing run and then freeze. Fixing the protocol up front is what keeps us from moving the goalposts after seeing the numbers. The criterion is a conservative non-overlap rule, which we prefer to a significance test because it makes no distributional assumptions: averaged over seeds, the pre-trained backbone must beat the random-initialisation baseline by more than the two error bars combined, so that the two seed distributions do not overlap at one standard deviation, that is $\mathrm{mean}_{\text{pre}}-\mathrm{std}_{\text{pre}} > \mathrm{mean}_{\text{rnd}}+\mathrm{std}_{\text{rnd}}$ on micro-mAP.

\subsection{Method}
\label{sec:probe-method}
\begin{figure}[t]
	\centering
	\includegraphics[width=\linewidth]{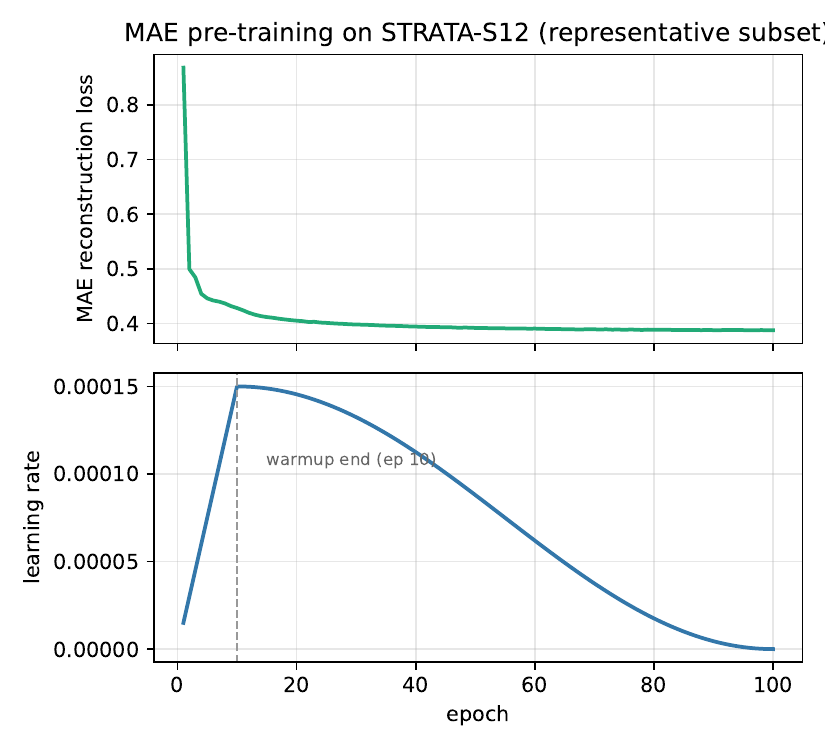}
	\caption{MAE pre-training on the representative subset of STRATA-S12: masked-reconstruction loss (top) and learning rate (bottom) over 100 epochs. The dashed line marks the end of the learning-rate warmup.}
	\label{fig:pretrain}
\end{figure}

We pre-train on a stratified representative subset of the corpus $D$ (\num{90403} pairs, the first 330 seed-shuffled WebDataset shards), sized to the pre-training I/O budget of the target hardware; because the shards are seed-shuffled at packing time, any prefix is a representative draw across all class$\times$season$\times$region strata. The backbone is a small vision transformer (ViT-S/16) trained as a masked autoencoder \citep{he2022mae}. Optical and radar are fused early, before the encoder: each pair becomes a single six-channel image --- the three Sentinel-2 RGB bands together with Sentinel-1 VV, VH and their difference --- so the model sees one co-registered stack rather than two separate streams. We mask 70\% of the patches and train the decoder to reconstruct the missing ones. The choice of early fusion and a single encoder is deliberate: the question here is not which fusion strategy works best, which belongs to the follow-up work, but whether the corpus supports useful pre-training at all, so we keep the setup plain and easy to reproduce while still using both modalities. Pre-training on the full corpus $D$ across the full model suite is deferred to the follow-up work, where compute is not the limiting factor. The full pre-training configuration is summarised in Table~\ref{tab:pretrain_config}, and pre-training converged as expected (Figure~\ref{fig:pretrain}): the masked-reconstruction loss fell from 0.87 to 0.39 over 100 epochs, with the learning rate following the warmup--cosine schedule (warmup over the first 10 epochs).

\input{pretrain/pretrain_config}

To read off how good the learned features are, we freeze the encoder and train only a linear classifier on top of it; linear probing of a frozen backbone is the standard way to read off representation quality in self-supervised learning~\citep{he2020moco,chen2020simclr}, and it avoids letting fine-tuning cover for a weak backbone. The downstream task is the refined BigEarthNet (reBEN / BigEarthNet~v2.0)~\citep{clasen2024reben} in its 19-class multi-label form~\citep{sumbul2021bigearthnetmm}, with micro-mAP as the headline number and macro-mAP reported beside it as a descriptor. The linear layer is trained on \ProbeMaxTrain\ training patches --- a seed-shuffled draw from the reBEN training split, fixed by seed \ProbeTrainSeed\ so the subset is reproducible --- and evaluated on the complete reBEN test split, so the score is read over the whole benchmark rather than a subsample of it. The baseline is the same architecture left at random initialisation and probed in exactly the same way, so the only difference between the two arms is whether the encoder was pre-trained on $D$. Each arm is run with at least three random seeds, and we report the mean and standard deviation. We keep to a single downstream task on purpose: the aim is to confirm that the corpus carries transferable signal, not to benchmark it, and a sweep across tasks, corpora and methods is left to the follow-up work.

Inputs are $224\times224$: native-resolution crops of the released $512\times512$ tiles at pre-training (centre crops at evaluation, with BigEarthNet patches resized up), standardised with fixed per-band statistics.

\subsection{Reproducibility layer}
The evaluation harness we release takes the corpus and the pre-trained weights as its only inputs and carries none of the acquisition-side code. The Phase-E removal ledger is released with the corpus: one row per removed pair, recording its stratum, whether it was removed by human review or by the automated post-download integrity check, and the Sentinel-1/Sentinel-2 source-scene identifiers, so any removal can be re-checked by re-fetching the scene. The evaluated encoder is the epoch-\ProbeEncoderEpoch\ checkpoint of the pre-training run of Section~\ref{sec:probe-method}, i.e.\ the fully trained backbone rather than an early snapshot; it is the weight file shipped with the release. Dependencies are pinned in a lockfile, seeds and splits are deterministic, and every run writes out its metrics, configuration and logs, so a reader can reproduce a number or trace where it came from. Pre-training was run on a single consumer RTX~4070 Laptop GPU (8~GB), which sets the pre-training I/O budget behind the representative subset of Section~\ref{sec:probe-method}; the recorded per-run hardware and configuration make the schedule reproducible on comparable devices.

\subsection{Results}
\begin{figure}[t]
	\centering
	\includegraphics[width=\linewidth]{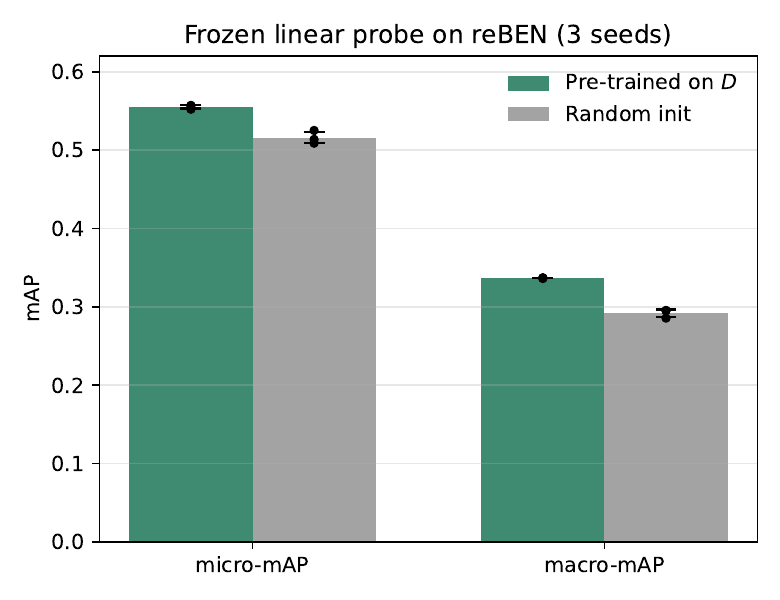}
	\caption{Frozen linear-probe scores on reBEN (BigEarthNet~v2.0), pre-trained
		on $D$ versus random initialisation. Bars are the mean over \ProbeSeeds{} seeds and the black dots the individual seed values; error bars are one standard deviation. Both micro-mAP (the primary metric of the pre-registered criterion) and macro-mAP are shown so that neither is read 	in isolation.}
	\label{fig:probe}
\end{figure}

The probe clears the pre-registered criterion, and by a comfortable margin. Table~\ref{tab:probe} and Figure~\ref{fig:probe} report the numbers behind that statement.

\begin{table}[t]
	\centering
	\caption{Frozen linear-probe performance on reBEN (BigEarthNet~v2.0, 19-class multi-label), mean $\pm$ standard deviation over \ProbeSeeds{} seeds. The pre-registered criterion is met when
		$\mathrm{mean}_{\text{pre}}-\mathrm{std}_{\text{pre}}$ exceeds
		$\mathrm{mean}_{\text{rnd}}+\mathrm{std}_{\text{rnd}}$ on micro-mAP.}
	\label{tab:probe}
	\begin{tabular}{lcc}
		\toprule
		Metric & Pre-trained on $D$ & Random init \\
		\midrule
		micro-mAP (primary)    & \ProbeMicroPreFull & \ProbeMicroRndFull \\
		macro-mAP (descriptor) & \ProbeMacroPreFull & \ProbeMacroRndFull \\
		\bottomrule
	\end{tabular}
\end{table}

The criterion is met. On micro-mAP the pre-trained backbone reaches \ProbeMicroPreFull\ against \ProbeMicroRndFull\ for the random-init baseline; subtracting one standard deviation from the first and adding one to the second still leaves the pre-trained arm ahead, by a margin of \ProbeCriterionMargin\ in micro-mAP, so the two seed distributions do not overlap at one sigma. The gap is modest in absolute terms, and we would rather state that than dress it up: the point of this check is a clean, pre-registered separation from random initialisation, not a headline score, and the separation is there with room to spare. The macro-mAP, which we report only as a descriptor, moves in the same direction and by a little more (\ProbeMacroPreFull\ versus \ProbeMacroRndFull, a gap of \ProbeMacroDelta), which is the expected pattern on a benchmark as class-imbalanced as BigEarthNet: macro weights the rare classes equally, so an encoder that has learnt something beyond the majority land covers gains more there than on the frequency-dominated micro average. Both arms are scored over all \ProbeClassesValid/\ProbeClassesTotal\ classes of the test split, with no class dropped for want of positives, so the two numbers are directly comparable.

\begin{figure}[t]
	\centering
	\includegraphics[width=\linewidth]{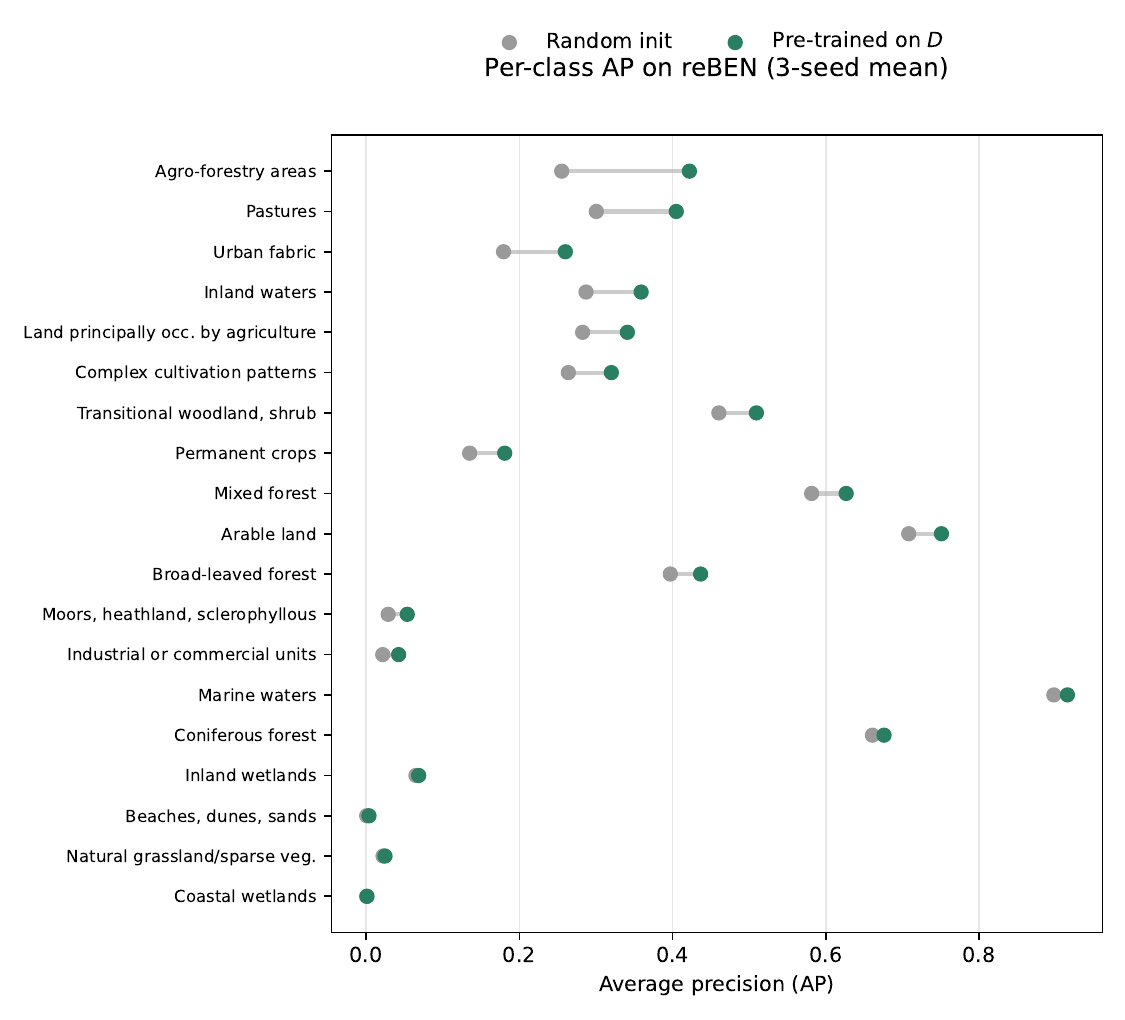}
	\caption{Per-class average precision on reBEN, pre-trained on $D$ versus random initialisation, averaged over \ProbeSeeds{} seeds. Each row is one of the \ProbeClassesTotal\ classes; the grey and green dots are the two arms and the connecting bar is the per-class gain from pre-training. Classes are ordered by that gain. The pre-trained encoder is ahead on every class, with the widest gains on the compositionally rich land covers and the narrowest on the classes either arm already separates well or that are too rare for a linear probe to recover.}
	\label{fig:perclass}
\end{figure}

Figure~\ref{fig:perclass} opens the macro number up by class. The pre-trained encoder is ahead of, or level with, the baseline on every one of the \ProbeClassesTotal\ classes --- there is no class where random initialisation wins --- and the size of the gain tracks how much structure a class carries.
The largest gains fall on the compositionally rich land covers, agro-forestry, pastures, urban fabric and complex cultivation, where separating the class needs texture and context rather than colour alone; these are exactly the ones a useful pre-training should help with. Where the two arms nearly coincide, the reason is visible in the other direction: marine water and coniferous forest sit high for both, because they are spectrally distinctive enough that even an untrained encoder ranks them well, so there is little headroom left to gain.
The three rarest classes --- coastal wetlands, beaches and dunes, natural grassland --- stay low for both arms; with on the order of a thousand positives across the whole of reBEN, a linear probe on frozen features does not recover them, pre-training or not. We flag this as a limit of the probe rather than of the corpus: it is a known consequence of BigEarthNet's class imbalance~\citep{clasen2024reben}, and a fuller evaluation with fine-tuning or class-balanced sampling is the place to revisit it, not this fitness-for-purpose check.

%% file: pretrain/pretrain_config.tex
\begin{table}[t]
\centering
\caption{MAE pre-training configuration (frozen for all seeds).}
\label{tab:pretrain_config}
\begin{tabular}{ll}
\toprule
Setting & Value \\
\midrule
Backbone & vit\_small\_patch16 \\
Input channels & 6 \\
Mask ratio & 0.7 \\
Effective batch size & 256 \\
Base learning rate & 1.5e-04 \\
Weight decay & 0.05 \\
Warmup epochs & 10 \\
Total epochs & 100 \\
Precision & bfloat16 \\
Seed & 0 \\
\bottomrule
\end{tabular}
\end{table}

%% file: sections/06_discussion.tex
\section{Discussion}
\label{sec:discussion}

\subsection{Usage Notes}
\label{sec:usage}

The expected primary use of the released corpus is SSL pre-training on the full corpus $D$, with downstream evaluation performed against external labelled benchmarks of the user's choosing. For users with restricted compute budgets, the reduced benchmark split $\dlite$ is the recommended drop-in: a small subset released purely for ease of use, preserving coverage of all {\AlgoStratK} strata at a bounded per-stratum size. It is capped, not equalised: a user who needs exactly equal strata should down-sample it further to its smallest stratum. The released pre-trained weights of Section~\ref{sec:validation} were obtained on a representative subset of $D$ (see Section~\ref{sec:probe-method}), calibrated to the pre-training budget of the target hardware.

Loaders should rely on the folder hierarchy for fast filtering and on the per-pair sidecar for any decision that depends on metadata (temporal gap, cloud fractions, label purity, provenance). The released README and Dataset Card mirror the conventions used here.

Users requiring supervised-task ground truth should rely on external labelled benchmarks; the WorldCover-derived class field of the corpus is intended for SSL stratification only, and the temporal gap between the WorldCover product (2021, v200) and the Sentinel acquisitions may introduce label noise. See the Label-source caveat in
Section~\ref{sec:labelcaveat}.

\subsection{Limitations}
\label{sec:limitations}
Several limitations should be stated plainly.
(i) The released corpus covers a broad set of regions but is not an exhaustive global acquisition. Running DDSA to full global completion is a matter of compute time rather than method; the contribution here is the acquisition algorithm and the evidence that its deficit-driven download improves stratum  balance, not a claim of planetary coverage. Extending the worked sample toward a complete global corpus is left as future work. The released corpus is a snapshot of an acquisition stopped by resource budget rather than by exhaustion of the frame: {\AlgoStratSparse} of the {\AlgoStratK} strata hold fewer than one hundred pairs, and those are the ones a continued run fills first.

(ii) Stratification labels come from ESA WorldCover 2021, four years before the imagery, so a small fraction of tiles in areas of real land-cover change may carry a stale stratum label	(Section~\ref{sec:labelcaveat}); the labels are used only as a stratification key, not as supervised targets.

(iii) The Phase~E quality control includes a human-in-the-loop step in which every accepted pair, including every recovered replacement, was inspected by a single reviewer. It is not fully automated and carries the usual reproducibility caveats of manual review, and a single reviewer leaves inter-rater agreement unmeasured; we mitigate this by fixing a closed criteria list (Section~\ref{sec:methods}), recording every manual removal under a single \texttt{manual\_review} category in $\Sigma$, and keeping full per-tile provenance so any removal can be re-fetched and re-checked. We do not log which criterion triggered each removal, as hand-labelling a small fraction by sub-type is disproportionate.
The step is retained because some failure modes have no per-pixel threshold signature --- cloud shadows, which are dark where the gate keys on bright pixels, and seamless temporal overlaps, whose two halves are each valid and only jointly inconsistent --- and therefore cannot be delegated to an automatic gate.

(iv) The recovery efficiency of Equation~\eqref{eq:recovery} is measured under DDSA's own quality gate and its bounded-retry policy, not against an absolute ceiling. A location counted as exhausted is one for which no scene passed \emph{these} thresholds within \emph{this} retry budget --- at most two retries per location, or fewer if the pool ran out; a laxer cloud threshold or a longer search would move some of those into the recovered column. The figure characterises the loop as configured here and is not a claim about the observability of the ground.
Retries were also allocated in stratum-deficit order rather than uniformly, so $\eta$ is an efficiency over the attempts the loop chose to make, not over a random sample of the failures.

(v) The technical validation is deliberately fit-for-purpose: a single method on a single downstream task, with pre-trained weights obtained on a representative subset of $D$ sized to the target hardware rather than on the full corpus. It confirms that the corpus carries transferable signal but does not benchmark it, nor test whether the full corpus or the broader stratum coverage improves representations; systematic cross-task, cross-corpus and cross-method comparison is left to follow-up work.

(vi) The coverage comparison against other corpora is descriptive rather than controlled, on two counts: the corpora were built over different sampling frames, and each is compared after its own quality filtering, which differs. \corpus applies its own multi-stage cascade, so part of any distributional difference could in principle follow from the filter rather than from the acquisition policy --- a persistently cloudy region loses more candidates under a strict gate than under a loose one. The deficit-driven ordering works against that mechanism rather than with it, since a stratum the gate depletes moves up the acquisition queue (Section~\ref{sec:phasec-ordering}), and the strongest regional differences appear in Africa and Oceania rather than in the cloudiest strata alone, but the two effects are not separated here. A controlled comparison would need a common sampling frame and a common gate across all three corpora, which is left to follow-up work.

(vii) The no-repair ablation shows the repair loop adding volume without measurably changing the distribution over the frame. We report this rather than the expected redistribution, and attribute the stratum balance to the deficit-driven planning of Phase~C instead. Whether a repair loop can be made to redistribute --- by prioritising under-filled strata over overall yield when choosing which failures to retry --- is a question this design does not answer.

(viii) The released tiles carry Sentinel-2 true colour and Sentinel-1 VV/VH rather than the full spectral and polarimetric stack --- a deliberate usability choice; the remaining bands are recoverable per tile from the recorded source identifiers (Section~\ref{sec:sidecar}).

(ix) The formal guarantee of Section~\ref{sec:formulation} covers the planning stage only: the greedy plan is $(1-1/e)$-optimal for the coverage objective, but the deficit-repair loop that acts on the realised (post-gate) coverage is characterised operationally --- as a bounded iteration measured by its recovery efficiency $\eta$ --- rather than with a coverage guarantee of its own. Establishing under what availability conditions the loop provably closes the residual deficit is left as future work.

%% file: sections/07_conclusions.tex
\section{Conclusions}
We presented DDSA, a deficit-driven stratified acquisition algorithm. Its planning stage solves a constrained maximum-coverage problem with the provably-optimal greedy rule (Section~\ref{sec:formulation}); its contribution is the online use of that plan --- a per-stratum deficit that, unlike opportunistic collection, is updated \emph{during} acquisition by every gate rejection, manual removal and recovery, and that the acquisition order follows. The algorithm is released together with \corpus, an open, fully documented SAR--optical corpus whose pairs carry STAC/PDS4-aligned, FAIR sidecars with checksum-verified integrity and full provenance. Because its frame follows land cover rather than population, the corpus places \num{26.7}\% of its pairs in Africa and \num{32.7}\% in the tropical band, against \num{14.1}\% and \num{18.9}\% for the city-centred SSL4EO-S12 (Section~\ref{sec:coverage-comparison}).

Three results, set out in Section~\ref{sec:algorithm-evidence} and only recalled here, follow from measuring the algorithm on its own product. The stratum balance of the corpus is a property of the acquisition \emph{ordering} --- the greedy submodular plan of Section~\ref{sec:formulation} --- not of how far the run was taken: replayed over every attempt the run made, failures included, deficit-driven ordering populates all {\AlgoStratK} strata within {\SimFullDeficit}\% of the attempts, against {\SimFullRoundRobin}\% for uniform round-robin effort and {\SimFullRandom}\% for a random order, and stays the most even at every partial budget, at the price of a lower yield per attempt (Section~\ref{sec:phasec-ordering}). Whether a failed acquisition attempt can be recovered is poorly predicted by the type of defect: cloud rejections are recovered at $\eta={\AlgoEtaCloudsPct\%}$ against ${\AlgoEtaNodataPct\%}$ for \texttt{nodata}, a gap most parsimoniously explained by the persistence of cloud through a season, though our metadata support that reading rather than establish it causally. And the repair loop supplies {\AlgoRepairSharePct\%} of the released pairs at no measurable cost in quality --- comparable temporal pairing and land-cover purity to first-pass pairs, and \AlgoQCSurvivalPct\% surviving the subsequent manual review --- with the balance itself coming from the deficit-driven ordering upstream rather than from the loop.
We deliberately restrict the claims of this paper to fitness-for-purpose. A pre-registered frozen-linear-probe check confirms that a backbone pre-trained on the corpus carries transferable signal, clearing a random-initialisation baseline under a conservative non-overlap criterion; it is a single-method, single-task check, not a benchmark, and it does not test whether the broader stratum coverage improves downstream performance. The released corpus is also a snapshot of an acquisition stopped by resource budget rather than by exhaustion of the frame, with {\AlgoStratSparse} of the {\AlgoStratK} strata still below one hundred pairs. What DDSA contributes is not a claim of planetary coverage but a reusable route to a coverage-balanced, quality-gated multi-modal corpus whose balance is controlled by the acquisition policy itself.

That property is what makes \corpus a substrate for the questions this paper does not answer. Because the corpus controls its geographic and seasonal composition by construction, it is a controlled ground on which SAR--optical modality asymmetry can be measured without it being conflated with where the samples happen to come from: how much a multi-modal model relies on each sensor and where that reliance fails, how representations degrade under each modality's characteristic failure, and whether a reliability-conditioned pre-training objective can act on the measured asymmetry. A controlled cross-corpus comparison with geographically held-out evaluation --- the direct test of the benefit of coverage --- is the immediate next step, for which both the algorithm and the corpus are intended as the foundation.

%% file: sections/08_backmatter.tex
\section*{Code Availability}
The evaluation harness used in Section~\ref{sec:validation}, together with its reproducibility layer, is available at
\url{https://github.com/csanchez-research/strata-s12-eval-harness}, archived at release on Zenodo (\url{https://doi.org/10.5281/zenodo.22892250}). It takes the corpus and the pre-trained weights as its only inputs and carries none of the acquisition-side code. The acquisition-side implementation is not redistributed: it is tied to operational credentials and infrastructure rather than to the method itself, and the pseudocode of Algorithms~\ref{alg:ddsa1} and~\ref{alg:ddsa2}, with the specification in this paper, is detailed enough to reimplement DDSA independently.

\section*{Data Availability}
The reduced split $\dlite$ of the \corpus corpus is openly available on Zenodo (\url{https://doi.org/10.5281/zenodo.22646276}) under a CC~BY~4.0 licence. It is distributed as eight per-class archives, each tile accompanied by a per-pair JSON sidecar, together with a manifest, per-archive checksums, a helper script and worked examples for immediate use. The full corpus $D$ is deposited on Zenodo (\url{https://doi.org/10.5281/zenodo.22699046}), which holds the metadata, the pre-trained SSL backbone weights of Section~\ref{sec:validation}, the per-stratum statistics report $\Sigma$, the post-acquisition integrity audit report, the Phase-E removal ledger (one row per pair withdrawn after acceptance, with its source-scene identifiers) and the Phase-D rejection ledger (one row per gate rejection, with its retry outcome, backing Table~\ref{tab:rejections}), and links to the image shards, which are hosted on Hugging Face (\url{https://huggingface.co/datasets/chsanleo/strata-s12}) owing to their size. All released artefacts are licensed CC~BY~4.0. The corpus is derived from Copernicus Sentinel-1 and Sentinel-2 data (\url{https://dataspace.copernicus.eu}) accessed through Google Earth Engine~\citep{gorelick2017gee}, and from ESA WorldCover 2021~v200~\citep{zanaga2022worldcover} and Natural Earth admin-0 boundaries~\citep{naturalearth}; each is redistributed under its own licence.

\section*{Acknowledgements}
This work contains modified Copernicus Sentinel data (2024–2025) processed by ESA, and uses ESA WorldCover 2021 v200 and Natural Earth data; the authors acknowledge the European Space Agency, the Copernicus Programme, and the WorldCover and Natural Earth teams for making these products openly available.

\section*{Funding}
This work was supported by the Key Technologies Research and Development Program of Liaoning Province under Grant No.~2024JH2/102400072, and by the Fundamental Research Funds for the Central Universities under Grant No.~N25BSS062.

\section*{CRediT authorship contribution statement}
\textbf{Christian Sanchez-Leon:} Conceptualization, Methodology, Software, Investigation, Data curation, Validation, Visualization, Writing -- original draft. \textbf{Dongming Chen:} Supervision, Funding acquisition, Writing -- review \& editing. \textbf{Dongqi Wang:} Writing -- review \& editing.

\section*{Declaration of competing interest}
The authors declare that they have no known competing financial interests or personal relationships that could have appeared to influence the work reported in this paper.

\section*{Declaration of generative AI and AI-assisted technologies in the manuscript preparation process}
During the preparation of this work the authors used Claude (Anthropic) for language editing and LaTeX formatting. After using this tool, the authors reviewed and edited the content as needed and take full responsibility for the content of the article.

%% file: main.bbl
\begin{thebibliography}{37}
\expandafter\ifx\csname natexlab\endcsname\relax\def\natexlab#1{#1}\fi
\providecommand{\url}[1]{\texttt{#1}}
\providecommand{\href}[2]{#2}
\providecommand{\path}[1]{#1}
\providecommand{\DOIprefix}{doi:}
\providecommand{\ArXivprefix}{arXiv:}
\providecommand{\URLprefix}{URL: }
\providecommand{\Pubmedprefix}{pmid:}
\providecommand{\doi}[1]{\href{http://dx.doi.org/#1}{\path{#1}}}
\providecommand{\Pubmed}[1]{\href{pmid:#1}{\path{#1}}}
\providecommand{\bibinfo}[2]{#2}
\ifx\xfnm\relax \def\xfnm[#1]{\unskip,\space#1}\fi
\bibitem[{Bastani et~al.(2023)Bastani, Wolters, Gupta, Ferdinando and
  Kembhavi}]{bastani2023satlas}
\bibinfo{author}{Bastani, F.}, \bibinfo{author}{Wolters, P.},
  \bibinfo{author}{Gupta, R.}, \bibinfo{author}{Ferdinando, J.},
  \bibinfo{author}{Kembhavi, A.}, \bibinfo{year}{2023}.
\newblock \bibinfo{title}{{SatlasPretrain}: A large-scale dataset for remote
  sensing image understanding}, in: \bibinfo{booktitle}{Proc. IEEE/CVF
  International Conference on Computer Vision (ICCV)}, pp.
  \bibinfo{pages}{16772--16782}.
\newblock \DOIprefix\doi{10.1109/ICCV51070.2023.01538}.
\bibitem[{Chen et~al.(2020)Chen, Kornblith, Norouzi and
  Hinton}]{chen2020simclr}
\bibinfo{author}{Chen, T.}, \bibinfo{author}{Kornblith, S.},
  \bibinfo{author}{Norouzi, M.}, \bibinfo{author}{Hinton, G.},
  \bibinfo{year}{2020}.
\newblock \bibinfo{title}{A simple framework for contrastive learning of visual
  representations}, in: \bibinfo{booktitle}{Proc. 37th International Conference
  on Machine Learning (ICML)}, pp. \bibinfo{pages}{1597--1607}.
\bibitem[{Clasen et~al.(2024)Clasen, Hackel, Burgert, Sumbul, Demir and
  Markl}]{clasen2024reben}
\bibinfo{author}{Clasen, K.N.}, \bibinfo{author}{Hackel, L.},
  \bibinfo{author}{Burgert, T.}, \bibinfo{author}{Sumbul, G.},
  \bibinfo{author}{Demir, B.}, \bibinfo{author}{Markl, V.},
  \bibinfo{year}{2024}.
\newblock \bibinfo{title}{{reBEN}: Refined {BigEarthNet} dataset for remote
  sensing image analysis}.
\newblock \bibinfo{note}{ArXiv:2407.03653}.
\bibitem[{Drusch et~al.(2012)Drusch, Del~Bello, Carlier, Colin, Fernandez,
  Gascon et~al.}]{drusch2012sentinel2}
\bibinfo{author}{Drusch, M.}, \bibinfo{author}{Del~Bello, U.},
  \bibinfo{author}{Carlier, S.}, \bibinfo{author}{Colin, O.},
  \bibinfo{author}{Fernandez, V.}, \bibinfo{author}{Gascon, F.}, et~al.,
  \bibinfo{year}{2012}.
\newblock \bibinfo{title}{{Sentinel-2}: {ESA}'s optical high-resolution mission
  for {GMES} operational services}.
\newblock \bibinfo{journal}{Remote Sensing of Environment}
  \bibinfo{volume}{120}, \bibinfo{pages}{25--36}.
\newblock \DOIprefix\doi{10.1016/j.rse.2011.11.026}.
\bibitem[{Feige(1998)}]{feige1998threshold}
\bibinfo{author}{Feige, U.}, \bibinfo{year}{1998}.
\newblock \bibinfo{title}{A threshold of $\ln n$ for approximating set cover}.
\newblock \bibinfo{journal}{Journal of the ACM} \bibinfo{volume}{45},
  \bibinfo{pages}{634--652}.
\newblock \DOIprefix\doi{10.1145/285055.285059}.
\bibitem[{Forgaard et~al.(2026)Forgaard, Reksten, Waldeland, Marsocci,
  Long{\'e}p{\'e}, Kampffmeyer and Salberg}]{thor2026}
\bibinfo{author}{Forgaard, T.}, \bibinfo{author}{Reksten, J.H.},
  \bibinfo{author}{Waldeland, A.U.}, \bibinfo{author}{Marsocci, V.},
  \bibinfo{author}{Long{\'e}p{\'e}, N.}, \bibinfo{author}{Kampffmeyer, M.},
  \bibinfo{author}{Salberg, A.B.}, \bibinfo{year}{2026}.
\newblock \bibinfo{title}{{THOR}: A versatile foundation model for earth
  observation climate and society applications}.
\newblock \bibinfo{note}{ArXiv:2601.16011}.
\bibitem[{Francis and Czerkawski(2024)}]{majortom2024}
\bibinfo{author}{Francis, A.}, \bibinfo{author}{Czerkawski, M.},
  \bibinfo{year}{2024}.
\newblock \bibinfo{title}{{Major TOM}: Expandable datasets for earth
  observation}, in: \bibinfo{booktitle}{IGARSS 2024 -- IEEE International
  Geoscience and Remote Sensing Symposium}, pp. \bibinfo{pages}{2935--2940}.
\bibitem[{Fuller et~al.(2023)Fuller, Millard and Green}]{fuller2023croma}
\bibinfo{author}{Fuller, A.}, \bibinfo{author}{Millard, K.},
  \bibinfo{author}{Green, J.R.}, \bibinfo{year}{2023}.
\newblock \bibinfo{title}{{CROMA}: Remote sensing representations with
  contrastive radar--optical masked autoencoders}, in:
  \bibinfo{booktitle}{Advances in Neural Information Processing Systems
  (NeurIPS)}.
\bibitem[{Gorelick et~al.(2017)Gorelick, Hancher, Dixon, Ilyushchenko, Thau and
  Moore}]{gorelick2017gee}
\bibinfo{author}{Gorelick, N.}, \bibinfo{author}{Hancher, M.},
  \bibinfo{author}{Dixon, M.}, \bibinfo{author}{Ilyushchenko, S.},
  \bibinfo{author}{Thau, D.}, \bibinfo{author}{Moore, R.},
  \bibinfo{year}{2017}.
\newblock \bibinfo{title}{{Google Earth Engine}: Planetary-scale geospatial
  analysis for everyone}.
\newblock \bibinfo{journal}{Remote Sensing of Environment}
  \bibinfo{volume}{202}, \bibinfo{pages}{18--27}.
\newblock \DOIprefix\doi{10.1016/j.rse.2017.06.031}.
\bibitem[{Hasan et~al.(2026)Hasan, Hossain, Roy, Bhowmik, Patel, Singha,
  Chaudhuri, Khan and Banerjee}]{geomeld2026}
\bibinfo{author}{Hasan, M.}, \bibinfo{author}{Hossain, M.A.},
  \bibinfo{author}{Roy, S.}, \bibinfo{author}{Bhowmik, S.},
  \bibinfo{author}{Patel, A.V.}, \bibinfo{author}{Singha, M.},
  \bibinfo{author}{Chaudhuri, S.}, \bibinfo{author}{Khan, M.H.},
  \bibinfo{author}{Banerjee, B.}, \bibinfo{year}{2026}.
\newblock \bibinfo{title}{{GeoMeld}: Toward semantically grounded foundation
  models for remote sensing}.
\newblock \bibinfo{note}{ArXiv:2604.10591}.
\bibitem[{He et~al.(2022)He, Chen, Xie, Li, Doll{\'a}r and
  Girshick}]{he2022mae}
\bibinfo{author}{He, K.}, \bibinfo{author}{Chen, X.}, \bibinfo{author}{Xie,
  S.}, \bibinfo{author}{Li, Y.}, \bibinfo{author}{Doll{\'a}r, P.},
  \bibinfo{author}{Girshick, R.}, \bibinfo{year}{2022}.
\newblock \bibinfo{title}{Masked autoencoders are scalable vision learners},
  in: \bibinfo{booktitle}{Proc. IEEE/CVF Conf. on Computer Vision and Pattern
  Recognition (CVPR)}.
\bibitem[{He et~al.(2020)He, Fan, Wu, Xie and Girshick}]{he2020moco}
\bibinfo{author}{He, K.}, \bibinfo{author}{Fan, H.}, \bibinfo{author}{Wu, Y.},
  \bibinfo{author}{Xie, S.}, \bibinfo{author}{Girshick, R.},
  \bibinfo{year}{2020}.
\newblock \bibinfo{title}{Momentum contrast for unsupervised visual
  representation learning}, in: \bibinfo{booktitle}{Proc. IEEE/CVF Conf. on
  Computer Vision and Pattern Recognition (CVPR)}, pp.
  \bibinfo{pages}{9729--9738}.
\newblock \DOIprefix\doi{10.1109/CVPR42600.2020.00975}.
\bibitem[{Justice et~al.(2002)Justice, Townshend, Vermote, Masuoka, Wolfe,
  Saleous, Roy and Morisette}]{justice2002modis}
\bibinfo{author}{Justice, C.O.}, \bibinfo{author}{Townshend, J.R.G.},
  \bibinfo{author}{Vermote, E.F.}, \bibinfo{author}{Masuoka, E.},
  \bibinfo{author}{Wolfe, R.E.}, \bibinfo{author}{Saleous, N.},
  \bibinfo{author}{Roy, D.P.}, \bibinfo{author}{Morisette, J.T.},
  \bibinfo{year}{2002}.
\newblock \bibinfo{title}{An overview of {MODIS} land data processing and
  product status}.
\newblock \bibinfo{journal}{Remote Sensing of Environment}
  \bibinfo{volume}{83}, \bibinfo{pages}{3--15}.
\newblock \DOIprefix\doi{10.1016/S0034-4257(02)00084-6}.
\bibitem[{Main-Knorn et~al.(2017)Main-Knorn, Pflug, Louis, Debaecker,
  M{\"u}ller-Wilm and Gascon}]{mainknorn2017sen2cor}
\bibinfo{author}{Main-Knorn, M.}, \bibinfo{author}{Pflug, B.},
  \bibinfo{author}{Louis, J.}, \bibinfo{author}{Debaecker, V.},
  \bibinfo{author}{M{\"u}ller-Wilm, U.}, \bibinfo{author}{Gascon, F.},
  \bibinfo{year}{2017}.
\newblock \bibinfo{title}{{Sen2Cor} for {Sentinel-2}}, in:
  \bibinfo{booktitle}{Image and Signal Processing for Remote Sensing XXIII}, p.
  \bibinfo{pages}{1042704}.
\newblock \DOIprefix\doi{10.1117/12.2278218}.
\bibitem[{{NASA Planetary Data System}(2024)}]{nasapds4}
\bibinfo{author}{{NASA Planetary Data System}}, \bibinfo{year}{2024}.
\newblock \bibinfo{title}{{PDS4 Information Model}}.
\newblock \bibinfo{howpublished}{\url{https://pds.nasa.gov/datastandards/}}.
\bibitem[{{Natural Earth}(2024)}]{naturalearth}
\bibinfo{author}{{Natural Earth}}, \bibinfo{year}{2024}.
\newblock \bibinfo{title}{Natural earth -- free vector and raster map data}.
\newblock \bibinfo{howpublished}{\url{https://www.naturalearthdata.com}}.
\bibitem[{Nedungadi et~al.(2024)Nedungadi, Kariryaa, Oehmcke, Belongie, Igel
  and Lang}]{nedungadi2024mmearth}
\bibinfo{author}{Nedungadi, V.}, \bibinfo{author}{Kariryaa, A.},
  \bibinfo{author}{Oehmcke, S.}, \bibinfo{author}{Belongie, S.},
  \bibinfo{author}{Igel, C.}, \bibinfo{author}{Lang, N.}, \bibinfo{year}{2024}.
\newblock \bibinfo{title}{{MMEarth}: Exploring multi-modal pretext tasks for
  geospatial representation learning}, in: \bibinfo{booktitle}{Computer Vision
  -- ECCV 2024}.
\newblock \DOIprefix\doi{10.1007/978-3-031-73039-9\_10}.
\bibitem[{Nemhauser et~al.(1978)Nemhauser, Wolsey and
  Fisher}]{nemhauser1978submodular}
\bibinfo{author}{Nemhauser, G.L.}, \bibinfo{author}{Wolsey, L.A.},
  \bibinfo{author}{Fisher, M.L.}, \bibinfo{year}{1978}.
\newblock \bibinfo{title}{An analysis of approximations for maximizing
  submodular set functions---{I}}.
\newblock \bibinfo{journal}{Mathematical Programming} \bibinfo{volume}{14},
  \bibinfo{pages}{265--294}.
\newblock \DOIprefix\doi{10.1007/BF01588971}.
\bibitem[{Olofsson et~al.(2014)Olofsson, Foody, Herold, Stehman, Woodcock and
  Wulder}]{olofsson2014good}
\bibinfo{author}{Olofsson, P.}, \bibinfo{author}{Foody, G.M.},
  \bibinfo{author}{Herold, M.}, \bibinfo{author}{Stehman, S.V.},
  \bibinfo{author}{Woodcock, C.E.}, \bibinfo{author}{Wulder, M.A.},
  \bibinfo{year}{2014}.
\newblock \bibinfo{title}{Good practices for estimating area and assessing
  accuracy of land change}.
\newblock \bibinfo{journal}{Remote Sensing of Environment}
  \bibinfo{volume}{148}, \bibinfo{pages}{42--57}.
\newblock \DOIprefix\doi{10.1016/j.rse.2014.02.015}.
\bibitem[{Purohit et~al.(2025)Purohit, Muhawenayo, Rolf and
  Kerner}]{spatialdist2025}
\bibinfo{author}{Purohit, M.}, \bibinfo{author}{Muhawenayo, G.},
  \bibinfo{author}{Rolf, E.}, \bibinfo{author}{Kerner, H.},
  \bibinfo{year}{2025}.
\newblock \bibinfo{title}{How does the spatial distribution of pre-training
  data affect geospatial foundation models?}
\newblock \bibinfo{note}{ArXiv:2501.12535; Good Data for Generative AI
  Workshop, AAAI 2025}.
\bibitem[{Schmitt et~al.(2019)Schmitt, Hughes, Qiu and
  Zhu}]{schmitt2019sen12ms}
\bibinfo{author}{Schmitt, M.}, \bibinfo{author}{Hughes, L.H.},
  \bibinfo{author}{Qiu, C.}, \bibinfo{author}{Zhu, X.X.}, \bibinfo{year}{2019}.
\newblock \bibinfo{title}{{SEN12MS} -- a curated dataset of georeferenced
  multi-spectral {Sentinel-1/2} imagery for deep learning and data fusion}, in:
  \bibinfo{booktitle}{ISPRS Annals of the Photogrammetry, Remote Sensing and
  Spatial Information Sciences}, pp. \bibinfo{pages}{153--160}.
\newblock \DOIprefix\doi{10.5194/isprs-annals-IV-2-W7-153-2019}.
\bibitem[{Snyder(1987)}]{snyder1987map}
\bibinfo{author}{Snyder, J.P.}, \bibinfo{year}{1987}.
\newblock \bibinfo{title}{Map Projections: A Working Manual}.
\newblock U.S. Geological Survey Professional Paper 1395,
  \bibinfo{publisher}{U.S. Government Printing Office}.
\bibitem[{{STAC Contributors}(2021)}]{stac2021spec}
\bibinfo{author}{{STAC Contributors}}, \bibinfo{year}{2021}.
\newblock \bibinfo{title}{{SpatioTemporal Asset Catalog (STAC) Specification
  v1.0.0}}.
\newblock \bibinfo{howpublished}{\url{https://stacspec.org}}.
\bibitem[{Stehman(2009)}]{stehman2009sampling}
\bibinfo{author}{Stehman, S.V.}, \bibinfo{year}{2009}.
\newblock \bibinfo{title}{Sampling designs for accuracy assessment of land
  cover}.
\newblock \bibinfo{journal}{International Journal of Remote Sensing}
  \bibinfo{volume}{30}, \bibinfo{pages}{5243--5272}.
\newblock \DOIprefix\doi{10.1080/01431160903131000}.
\bibitem[{Sumbul et~al.(2019)Sumbul, Charfuelan, Demir and
  Markl}]{sumbul2019bigearthnet}
\bibinfo{author}{Sumbul, G.}, \bibinfo{author}{Charfuelan, M.},
  \bibinfo{author}{Demir, B.}, \bibinfo{author}{Markl, V.},
  \bibinfo{year}{2019}.
\newblock \bibinfo{title}{{BigEarthNet}: A large-scale benchmark archive for
  remote sensing image understanding}, in: \bibinfo{booktitle}{IGARSS 2019 --
  IEEE International Geoscience and Remote Sensing Symposium}, pp.
  \bibinfo{pages}{5901--5904}.
\newblock \DOIprefix\doi{10.1109/IGARSS.2019.8900532}.
\bibitem[{Sumbul et~al.(2021)Sumbul, de~Wall, Kreuziger, Marcelino, Costa,
  Benevides, Caetano, Demir and Markl}]{sumbul2021bigearthnetmm}
\bibinfo{author}{Sumbul, G.}, \bibinfo{author}{de~Wall, A.},
  \bibinfo{author}{Kreuziger, T.}, \bibinfo{author}{Marcelino, F.},
  \bibinfo{author}{Costa, H.}, \bibinfo{author}{Benevides, P.},
  \bibinfo{author}{Caetano, M.}, \bibinfo{author}{Demir, B.},
  \bibinfo{author}{Markl, V.}, \bibinfo{year}{2021}.
\newblock \bibinfo{title}{{BigEarthNet-MM}: A large-scale, multimodal,
  multilabel benchmark archive for remote sensing image classification and
  retrieval}.
\newblock \bibinfo{journal}{IEEE Geoscience and Remote Sensing Magazine}
  \bibinfo{volume}{9}, \bibinfo{pages}{174--180}.
\newblock \DOIprefix\doi{10.1109/MGRS.2021.3089174}.
\bibitem[{Torralba and Efros(2011)}]{torralba2011bias}
\bibinfo{author}{Torralba, A.}, \bibinfo{author}{Efros, A.A.},
  \bibinfo{year}{2011}.
\newblock \bibinfo{title}{Unbiased look at dataset bias}, in:
  \bibinfo{booktitle}{CVPR 2011 -- IEEE Conference on Computer Vision and
  Pattern Recognition}, pp. \bibinfo{pages}{1521--1528}.
\newblock \DOIprefix\doi{10.1109/CVPR.2011.5995347}.
\bibitem[{Torres et~al.(2012)Torres, Snoeij, Geudtner, Bibby, Davidson, Attema
  et~al.}]{torres2012sentinel1}
\bibinfo{author}{Torres, R.}, \bibinfo{author}{Snoeij, P.},
  \bibinfo{author}{Geudtner, D.}, \bibinfo{author}{Bibby, D.},
  \bibinfo{author}{Davidson, M.}, \bibinfo{author}{Attema, E.}, et~al.,
  \bibinfo{year}{2012}.
\newblock \bibinfo{title}{{GMES Sentinel-1} mission}.
\newblock \bibinfo{journal}{Remote Sensing of Environment}
  \bibinfo{volume}{120}, \bibinfo{pages}{9--24}.
\newblock \DOIprefix\doi{10.1016/j.rse.2011.05.028}.
\bibitem[{Tseng et~al.(2023)Tseng, Cartuyvels, Zvonkov, Purohit, Rolnick and
  Kerner}]{presto2023}
\bibinfo{author}{Tseng, G.}, \bibinfo{author}{Cartuyvels, R.},
  \bibinfo{author}{Zvonkov, I.}, \bibinfo{author}{Purohit, M.},
  \bibinfo{author}{Rolnick, D.}, \bibinfo{author}{Kerner, H.},
  \bibinfo{year}{2023}.
\newblock \bibinfo{title}{Lightweight, pre-trained transformers for remote
  sensing timeseries}.
\newblock \bibinfo{note}{ArXiv:2304.14065}.
\bibitem[{Tuia et~al.(2016)Tuia, Persello and
  Bruzzone}]{tuia2016domainadaptation}
\bibinfo{author}{Tuia, D.}, \bibinfo{author}{Persello, C.},
  \bibinfo{author}{Bruzzone, L.}, \bibinfo{year}{2016}.
\newblock \bibinfo{title}{Domain adaptation for the classification of remote
  sensing data: An overview of recent advances}.
\newblock \bibinfo{journal}{IEEE Geoscience and Remote Sensing Magazine}
  \bibinfo{volume}{4}, \bibinfo{pages}{41--57}.
\newblock \DOIprefix\doi{10.1109/MGRS.2016.2548504}.
\bibitem[{Wang et~al.(2024)Wang, Albrecht, Braham, Liu, Xiong and
  Zhu}]{wang2024decur}
\bibinfo{author}{Wang, Y.}, \bibinfo{author}{Albrecht, C.M.},
  \bibinfo{author}{Braham, N.A.A.}, \bibinfo{author}{Liu, C.},
  \bibinfo{author}{Xiong, Z.}, \bibinfo{author}{Zhu, X.X.},
  \bibinfo{year}{2024}.
\newblock \bibinfo{title}{Decoupling common and unique representations for
  multimodal self-supervised learning}, in: \bibinfo{booktitle}{Computer Vision
  -- ECCV 2024}, pp. \bibinfo{pages}{286--303}.
\newblock \DOIprefix\doi{10.1007/978-3-031-73397-0\_17}.
\bibitem[{Wang et~al.(2022)Wang, Albrecht, Braham, Mou and
  Zhu}]{wang2022sslreview}
\bibinfo{author}{Wang, Y.}, \bibinfo{author}{Albrecht, C.M.},
  \bibinfo{author}{Braham, N.A.A.}, \bibinfo{author}{Mou, L.},
  \bibinfo{author}{Zhu, X.X.}, \bibinfo{year}{2022}.
\newblock \bibinfo{title}{Self-supervised learning in remote sensing: A
  review}.
\newblock \bibinfo{journal}{IEEE Geoscience and Remote Sensing Magazine}
  \bibinfo{volume}{10}, \bibinfo{pages}{213--247}.
\newblock \DOIprefix\doi{10.1109/MGRS.2022.3198244}.
\bibitem[{Wang et~al.(2023)Wang, Braham, Xiong, Liu, Albrecht and
  Zhu}]{wang2023ssl4eo}
\bibinfo{author}{Wang, Y.}, \bibinfo{author}{Braham, N.A.A.},
  \bibinfo{author}{Xiong, Z.}, \bibinfo{author}{Liu, C.},
  \bibinfo{author}{Albrecht, C.M.}, \bibinfo{author}{Zhu, X.X.},
  \bibinfo{year}{2023}.
\newblock \bibinfo{title}{{SSL4EO-S12}: A large-scale multimodal, multitemporal
  dataset for self-supervised learning in {Earth} observation}.
\newblock \bibinfo{journal}{IEEE Geoscience and Remote Sensing Magazine}
  \bibinfo{volume}{11}, \bibinfo{pages}{98--106}.
\newblock \DOIprefix\doi{10.1109/MGRS.2023.3281651}.
\bibitem[{Wickrema et~al.(2025)Wickrema, Mace, Brown, Cabrera, Krall, O'Neill,
  Sarkar, Weissman, Hughes and Zarrella}]{akupara2025}
\bibinfo{author}{Wickrema, C.}, \bibinfo{author}{Mace, E.},
  \bibinfo{author}{Brown, H.}, \bibinfo{author}{Cabrera, H.},
  \bibinfo{author}{Krall, N.}, \bibinfo{author}{O'Neill, M.},
  \bibinfo{author}{Sarkar, S.}, \bibinfo{author}{Weissman, L.},
  \bibinfo{author}{Hughes, E.}, \bibinfo{author}{Zarrella, G.},
  \bibinfo{year}{2025}.
\newblock \bibinfo{title}{Scaling remote sensing foundation models: Data domain
  tradeoffs at the peta-scale}.
\newblock \bibinfo{note}{ArXiv:2512.23903}.
\bibitem[{Wilkinson et~al.(2016)}]{wilkinson2016fair}
\bibinfo{author}{Wilkinson, M.D.}, et~al., \bibinfo{year}{2016}.
\newblock \bibinfo{title}{The {FAIR} guiding principles for scientific data
  management and stewardship}.
\newblock \bibinfo{journal}{Scientific Data} \bibinfo{volume}{3},
  \bibinfo{pages}{160018}.
\newblock \DOIprefix\doi{10.1038/sdata.2016.18}.
\bibitem[{Wilson and Jetz(2016)}]{wilson2016cloud}
\bibinfo{author}{Wilson, A.M.}, \bibinfo{author}{Jetz, W.},
  \bibinfo{year}{2016}.
\newblock \bibinfo{title}{Remotely sensed high-resolution global cloud dynamics
  for predicting ecosystem and biodiversity distributions}.
\newblock \bibinfo{journal}{PLOS Biology} \bibinfo{volume}{14},
  \bibinfo{pages}{e1002415}.
\newblock \DOIprefix\doi{10.1371/journal.pbio.1002415}.
\bibitem[{Zanaga et~al.(2022)Zanaga, Van De~Kerchove, Daems, De~Keersmaecker,
  Brockmann, Kirches, Wevers, Cartus, Santoro, Fritz, Lesiv, Herold,
  Tsendbazar, Xu, Ramoino and Arino}]{zanaga2022worldcover}
\bibinfo{author}{Zanaga, D.}, \bibinfo{author}{Van De~Kerchove, R.},
  \bibinfo{author}{Daems, D.}, \bibinfo{author}{De~Keersmaecker, W.},
  \bibinfo{author}{Brockmann, C.}, \bibinfo{author}{Kirches, G.},
  \bibinfo{author}{Wevers, J.}, \bibinfo{author}{Cartus, O.},
  \bibinfo{author}{Santoro, M.}, \bibinfo{author}{Fritz, S.},
  \bibinfo{author}{Lesiv, M.}, \bibinfo{author}{Herold, M.},
  \bibinfo{author}{Tsendbazar, N.E.}, \bibinfo{author}{Xu, P.},
  \bibinfo{author}{Ramoino, F.}, \bibinfo{author}{Arino, O.},
  \bibinfo{year}{2022}.
\newblock \bibinfo{title}{{ESA WorldCover 10 m 2021 v200}}.
\newblock \bibinfo{howpublished}{Zenodo}.
\newblock \DOIprefix\doi{10.5281/zenodo.7254221}.

\end{thebibliography}
